\documentclass[aps,prl,amsmath,floatfix]{revtex4}
\usepackage{graphicx}
\usepackage{color}
\usepackage{bm}
\usepackage{epsfig}
\usepackage{latexsym}
\usepackage{pifont}
\usepackage{float}
\usepackage[utf8]{inputenc}
\graphicspath{{figure/}}
\usepackage{siunitx}
\usepackage{textgreek}

\begin{document}
\newcommand{\be}{\begin{equation}}
\newcommand{\ee}{\end{equation}}
\newcommand{\bea}{\begin{eqnarray}}
\newcommand{\eea}{\end{eqnarray}}
\newcommand{\nn}{\nonumber}
\renewcommand{\figurename}{Fig.}

\title{Chemotaxis of branched cells in complex environments}

\author{Jiayi Liu$^{1,2}$, Jonathan E. Ron$^{2}$, Giulia Rinaldi$^{3}$, Ivanna Williantarra$^{3}$, Antonios Georgantzoglou$^{3,4}$, Ingrid de Vries$^{5}$, Michael Sixt$^{5}$, Milka Sarris$^{3}$ and Nir S. Gov$^{2,3}$}
\affiliation{$^{1}$Department of Physics, Yale University, New Haven, CT, USA}
\affiliation{$^{2}$Department of Chemical and Biological Physics, Weizmann Institute of Science, Rehovot, Israel}
\affiliation{$^{3}$Department of Physiology, Development and Neuroscience, Downing Site, University of Cambridge, Cambridge, UK}
\affiliation{$^{4}$Novo Nordisk Foundation Center for Stem Cell Medicine (reNEW), Department of Biomedical Sciences, University of Copenhagen, Denmark}
\affiliation{$^{5}$Institute of Science and Technology Austria (ISTA), Klosterneuburg, Austria}

\begin{abstract}
Cell migration \textit{in vivo} is often guided by chemical signals. Such chemotaxis, such as performed by immune cells migrating to a wound site, is complicated by the complex geometry inside living tissues. In this study, we extend our theoretical model of branched-cell migration on a network by introducing chemokine sources to explore the cellular response. The model predicts a speed-accuracy tradeoff, whereby slow cells are significantly more accurate and able to follow efficiently a weak chemoattractant signal. We then compare the model's predictions with experimental observations of neutrophils migrating to the site of laser-inflicted wound in a zebrafish larva fin, and migrating \textit{in-vitro} inside a regular lattice of pillars. We find that the model captures the details of the sub-cellular response to the chemokine gradient, as well as the large-scale migration response. This comparison suggests that the neutrophils behave as fast cells, compromising their chemotaxis accuracy, which explains the functionality of these immune cells. 
\end{abstract}

\maketitle

\section*{Introduction}

\begin{figure}[h!]
    \centering
    \includegraphics[width=1\textwidth]{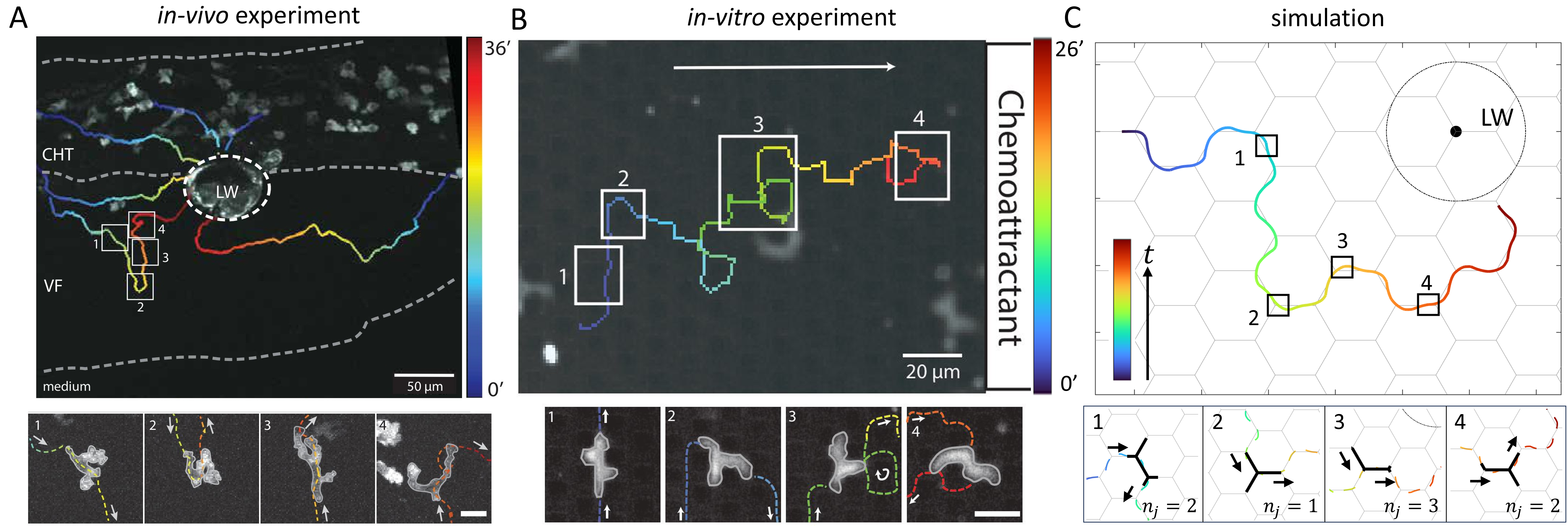}
    \caption{(A) Morphological dynamics of neutrophil swarming during migration from the caudal hematopoietic tissue (CHT) toward a laser-induced wound (LW) at the ventral fin (VF)–CHT boundary in transgenic zebrafish larva expressing the calcium indicator GCamp6f (see SI section S-1 for more details). Top panel: Representative trajectory plots of neutrophils migrating toward the LW, with trajectory color denoting migration time. White boxes highlight time points along the trajectory of an individual cell, corresponding to the images in the bottom panels (1–4). White arrows indicate the direction of migration. Scale bars: 50 $\mu$m for the trajectory plots and 10 $\mu$m for the enlarged images. (B) Trajectory and snapshots of the cell shape, while performing chemotaxis inside a regular lattice of pillars (see SI section S-2 for more details) \cite{renkawitz2018micro}. (C) Simulation of a branched cell moving toward a chemokine point source. Top panel: A representative trajectory of the cell's center of mass (C.O.M.) during migration, with trajectory color denoting migration time. Black boxes highlight time points along the trajectory of the cell, corresponding to the images in the bottom panels (1–4). Black arrows indicate the direction of migration. Key parameters: \(C/c_0=0.01, \epsilon=0.2, d=3, \beta_0=8, \sigma=0.8\).}
    \label{fig1}
\end{figure}

Tissues have complex geometries and present a great diversity of topologies which force embedded motile cells to deform and exhibit a plethora of branched shapes \cite{bodor2020cell}. Archetypal examples include immune cells that continuously scan the tissues in search for signals and need to migrate within the complex tissues towards sites of wound and infection \cite{saez2018leukocyte,vargas2017mechanisms,georgantzoglou2022two} (Fig.~\ref{fig1}), or cancer cells that spread from the main tumor \cite{friedl2011cancer}. In order to efficiently migrate towards their target, branched cells must coordinate their branch dynamics to navigate the microenvironment and decide on the new migration direction, in a process termed directional decision-making (DDM) \cite{garcia2019reconstitution,sarris2015navigating}. Despite its clear biological importance, the mechanisms of branched cell migration and chemotaxis through complex environments remain poorly understood \cite{yamada2019mechanisms}. Furthermore, there is currently a lack of theoretical framework for tackling these questions.

We have recently developed a theoretical model describing the shape dynamics of branched migrating cells performing DDM over single junctions \cite{ron2024emergent}, and the spontaneous polarization and migration of highly branched cells on hexagonal networks \cite{liu2024shape}. The coarse-grained model describes cells migrating on an hexagonal network, composed of linear segments (Fig.~\ref{fig1}C), with an internal mechanism for spontaneous cellular polariaztion and migration.

Here we utilize this theoretical model to explore the chemotaxis characteristics of branched cells. To test the simulations we compared them to two systems (Fig.~\ref{fig1}), one \textit{in vivo} using tissue injury in zebrafish \cite{poplimont2020neutrophil,georgantzoglou2022two} and one \textit{in-vitro} where cells migrate in a regular lattice of pillars \cite{renkawitz2018micro}. We validate the model by comparing to experimental data of the shape and migration dynamics of neutrophils as they respond to a signal that recruits them to a wound, in the zebra-fish skin \cite{georgantzoglou2022two}. Our model provides insights into the constraints that determine the optimal migration properties of neutrophils.

\section*{Chemotaxis over a single junction}
\begin{figure}[h!]
    \centering
    \includegraphics[width=1\textwidth]{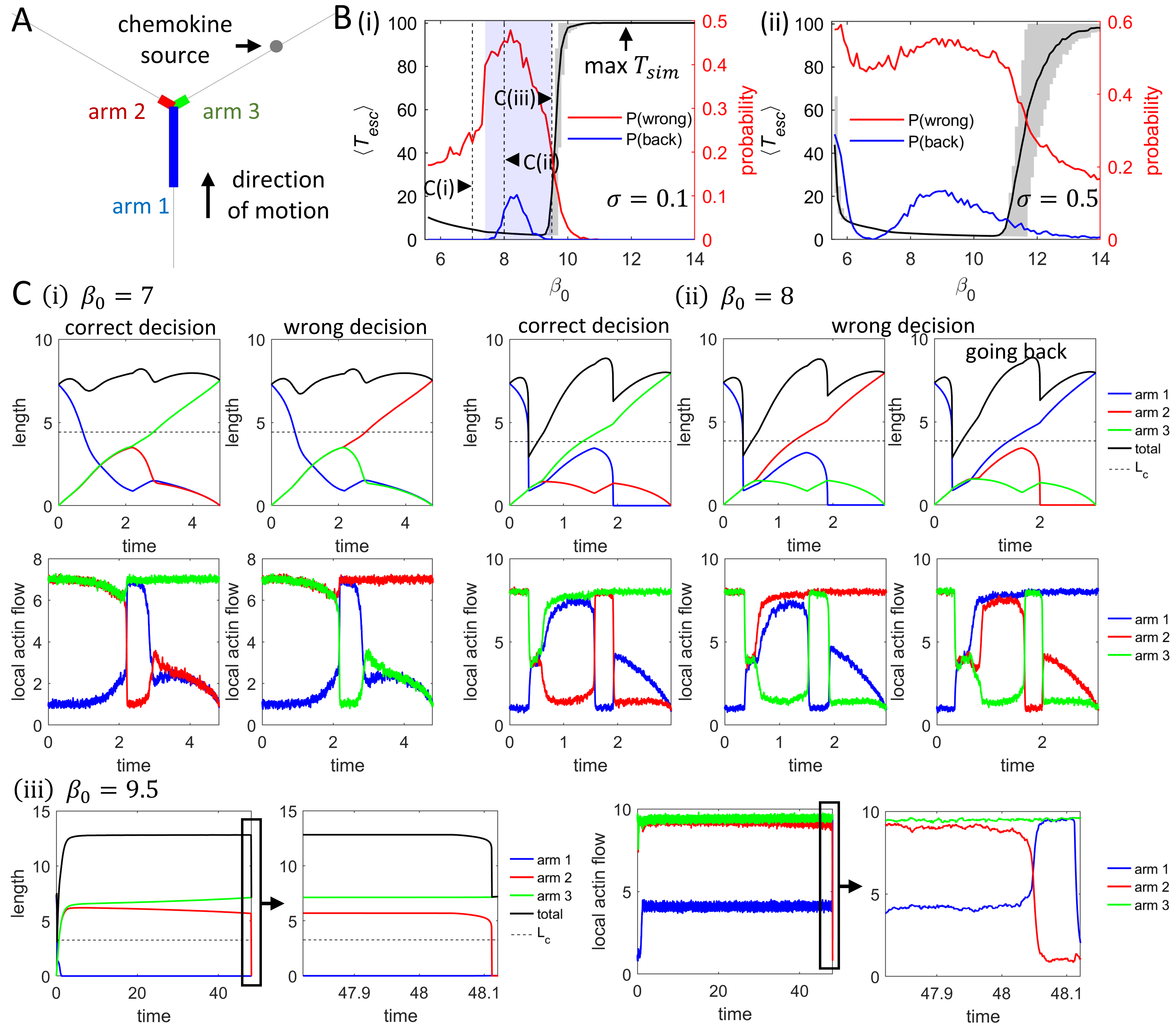}
    \caption{Chemotaxis dynamics over a single-junction with a localized chemokine source. (A) Schematic representation of the model. (B) Mean time for the cell to traverse the junction \(T_{esc}\) (black), mean error rate \(P(wrong)\) (the probability of leaving along arm 2, red), and the mean probability of being "reflected" at the junction \(P(back)\) (leaving along arm 1, blue), as functions of \(\beta_0\). Maximum simulation time: \(T=100\). The blue shaded region in (i) marks the range where cells undergo stick-slip events, and a cell can reverse its path (\(P(back)>0\)). Panels (i) and (ii) correspond to two values of the internal noise in the actin polymerization activity \(\sigma=0.1\) and \(\sigma=0.5\), respectively. (C) Time series of length of the arms and total cell length (top), as well as local actin flows at the tips of arms (bottom), for representative events: making the correct decision (leaving along arm 3), making the wrong decision while moving forward (leaving along arm 2), and making the wrong decision by turning back (leaving along arm 1). Panels (i), (ii), and (iii) correspond to \(\beta_0=7.0\), \(\beta_0=8.0\) and \(\beta_0=9.5\), respectively. Other key parameter: \(\epsilon=0.001\).}
    \label{fig2}
\end{figure}

We start by studying the chemotaxis of a cell when migrating over a single junction (Fig.~\ref{fig2}A). We extend our previous model \cite{ron2020one,ron2024emergent}, by introducing the effect of an external chemical signal (see Supplementary Information section S-3 for the model equations). The binding of the chemokine to the receptors at the leading edge of the cell triggers a local enhancement of the activity of the actin polymerization machinery \cite{insall2009actin}. This is represented in our model by an increase in the value of the parameter $\beta$, which describes the strength of the actin flow and protrusive force acting at the tips of the arms of the branched cell. For a single junction we simplify the signal by assuming that it is localized along one of the arms that extend from the junction (arm 3, Fig.~\ref{fig2}A), described as:

\begin{equation}
\beta = \beta_0 \, (1+\epsilon)
\label{eq1}
\end{equation}

\noindent
where \(\beta_0\) represents the actin polymerization speed parameter in the absence of external signals, and \(\epsilon\) is the increase factor due to the chemokine signal. For the tips of the other two arms (arms 1 and 2), the polymerization speed parameter remains unchanged, with \(\beta = \beta_0\). 

As in our previous work \cite{ron2024emergent}, the model equations (see SI section S-3) are normalized by the timescale of the inverse of the focal adhesion disassembly rate ($5–30$ min \cite{ivaska2012unanchoring,ren2000focal,berginski2011high,gupton2006spatiotemporal}) and by the length scale of the rest length of the cell ($10–100$ $\mu m$ on one-dimensional tracks \cite{ron2020one}). In the rest of this paper, we kept all the parameters of the cells as were calibrated to fit the observed dynamics of typical motile cells \cite{ron2024emergent} (Table. S-1). 

Our main free control parameter is the value of the actin polymerization (and retrograde flow) activity, $\beta_0$. This parameter has a lower bound given by the minimal value ($\beta_c$) that allows the cells to polarize \cite{ron2020one} and migrate over the network \cite{ron2024emergent}. This value is weakly dependent on the cell shape, increasing with the number of network nodes and branches that the cell has. To ensure that our model cell is able to migrate, we therefore explore the dynamics for $\beta_0>\beta_c$ (where $\beta_c\sim 5-6$ for the model parameters chosen here \cite{ron2024emergent}). Above a higher critical value, $\beta_c^{slow}$, the model predicts that the cell may get stuck in a "slow-mode" process, where the two leading arms are highly elongated while the back is stuck at the same node, until the competition between the two elongated arms is resolved \cite{ron2024emergent}. We find that $\beta_c^{slow}\sim9-12$ for the model parameters chosen here (Table. S-1). These considerations determine the range of $\beta_0$ values that we explore in this paper (see Fig.~\ref{fig2}B for example).

In Fig.~\ref{fig2}B(i,ii) we plotted for a fixed small bias $\epsilon=0.001$ (Eq.\ref{eq1}) and two different levels of the noise the mean escape time that it takes for the cell to migrate over the junction, \(\langle T_{esc} \rangle\) , and the probability of making the wrong decision, \(P(wrong)\), as functions of \(\beta_0\) (Fig.~\ref{fig2}B). The wrong decision indicates that the cell does not leave the junction along the path that leads to the chemokine source (arm 3), but rather leaves along either arm 2 or arm 1. The proportion of events where the cell is "reflected" by the junction, i.e. turn back and migrate along arm 1, are denoted by the blue line. The region where stick-slip migration over the junction occurs is denoted by the blue shading, and corresponds to the region where the reflection probability is nonzero. Typical examples of the cellular dynamics during different cases of directional decision-making are shown in Fig.~\ref{fig2}C (for two values of $\beta_0$ denoted by vertical dashed lines in Fig.~\ref{fig2}B(i)).

For weak noise (\(\sigma=0.1\), Fig.~\ref{fig2}B(i))), the mean escape time (\(\langle T_{esc} \rangle\)) decreases as $\beta_0$ increases, which is expected as the cell velocity increases with an increase of actin polymerization activity \cite{ron2024emergent}. At large values of $\beta_0\sim10$ the mean escape time sharply increases, as the cell tends to remain for a long period of time in the "slow-mode", with two highly elongated arms that are locked in symmetric competition (see Fig.~\ref{fig2}C(iii)), consistent with the cell behavior without chemotaxis \cite{ron2024emergent}. We see that in this mode the rear arm of the cell (arm 1) shrinks approximately to zero and the two new arms remain elongated with similar lengths for a long time, until one of the arms rapidly retracts.

The accuracy, as quantified by the mean error probability \(P(wrong)\), shows an opposite trend, exhibiting a “speed-accuracy tradeoff”. Specifically, cells that spend less time over the junction are more likely to make incorrect decisions and miss their way to the target. The abrupt increase in \(P(wrong)\) for small noise at \(\beta_0 \approx 8\) arises from the emergence of a finite probability for the cell to reflect along the original path (blue line in Fig.~\ref{fig2}B(i)). This behavior occurs exclusively during stick-slip events \cite{ron2020one, ron2024emergent}, which is observed only for larger \(\beta\) (in the blue-shaded regime of Fig.~\ref{fig2}B(i)). Examples of smooth and stick-slip migrations are shown in Fig.~(\ref{fig2}C(i,ii)). For smooth motion, the only possible incorrect decision is to leave the junction along the new path without the source (arm 2). However, since during stick-slip events the cell length decreases below the critical polarization length $L_c$ (denoted by the horizontal dashed line in Fig.~\ref{fig2}C(i,ii)), cells may also escape the junction by returning along their original path (arm 1). The duration that the cell spends with a length below $L_c$, and loses its polarity, causes the sharp increase in accuracy errors (similar to behavior at high noise and very low $\beta_0$, as discussed below).

As \(\beta_0\) increases further, the cell transitions into the "slow mode", characterized by two elongated and competing arms along the new paths, as observed in the one-junction model \cite{ron2024emergent} and shown in Fig.~\ref{fig2}C(iii), for two noise levels. We calculated the probability of slow mode occurrence, \(P(slow)\), as a function of \(\beta_0\) (Fig.~S-3A), and found that it closely aligns with the trend of the mean escape time (\(\langle T_{esc} \rangle\)) versus against the actin polymerization speed baseline (\(\beta_0\)) (Fig.~\ref{fig2}B(i)). Additionally, we computed the probability of making incorrect decisions in the fast and slow modes (red solid and dashed lines in Fig.~S3-A, respectively), showing that the slow mode exhibits significantly higher accuracy.

Finally, for specific values of \(\beta_0\) (denoted by black dashed lines in Fig.~S-3A), we plotted in Fig.~S-3B(i-iii) the distribution of escape times \(T_{esc}\) (histograms) and the value of \(P(wrong)\) as function of \(T_{esc}\) (black line). We observe a complex relation between the speed of individual events and their accuracy. This complex relation arises from two competing effects: slower events can be more accurate due to the speed-accuracy relation that appears on the average values. On the other hand, higher escape times can be associated with events of polarity loss, which give rise to higher error rates.

For larger noise (\(\sigma=0.5\)), we find a similar tradeoff as for the low noise, except at very low values of \(\beta_0\) (Fig.~\ref{fig2}B(ii)). In this limit, due to the large noise the cell loses its polarization when migrating over the junction, and has an almost equal probability of leaving along any of the three arms, leading to a $P(wrong)\sim2/3$. 

Taken together, these results confirm the existence of a tradeoff between speed and accuracy of cellular DDM during chemotaxis. In the next section we explore the consequences of this behavior for the large-scale chemotactic migration of cells over a hexagonal network.

\section*{Branched cell chemotaxis on a hexagonal network}

We next studied the migration dynamics of a branched cell on a hexagonal network, in the presence of a global chemokine gradient. 

Our treatment of a branched cell that spans more than one junction was described in \cite{liu2024shape}, and explained in more details in SI section S-3. Here, we expose these branched cells to a chemokine concentration field that has an exponential form:

\begin{equation}  \label{c(r)}
c(r) = c_0 \, \exp\left(-\frac{r}{r_0}\right)
\end{equation}

\noindent
where \(c_0\) is the chemokine concentration at the source position, \(r_0\) is the decay length of the chemokine concentration, and \(r\) is the distance from the tips of the cellular arms to the chemokine source. The local actin polymerization activity at each arm tip is enhanced due to the chemokine concentration profile given by (similar to Eq.\ref{eq1}):

\begin{equation} \label{b(r)}
\beta(r) = \beta_0 \, \left[1 + \epsilon \, \left(1+\frac{C}{c_0}\right) \, \frac{c(r)}{c(r)+C}\right]
\end{equation}

\noindent
where \(\beta_0\) is the actin polymerization speed in the absence of external signals, \(\epsilon\) is the maximum enhancement of \(\beta\), and \(C\) is the saturation concentration for detecting the signal. The dimensionless factor \((1+C/c_0)\) normalizes the enhancement to be maximal at \(r=0\).

\begin{figure}[h!]
    \centering
    \includegraphics[width=1\textwidth]{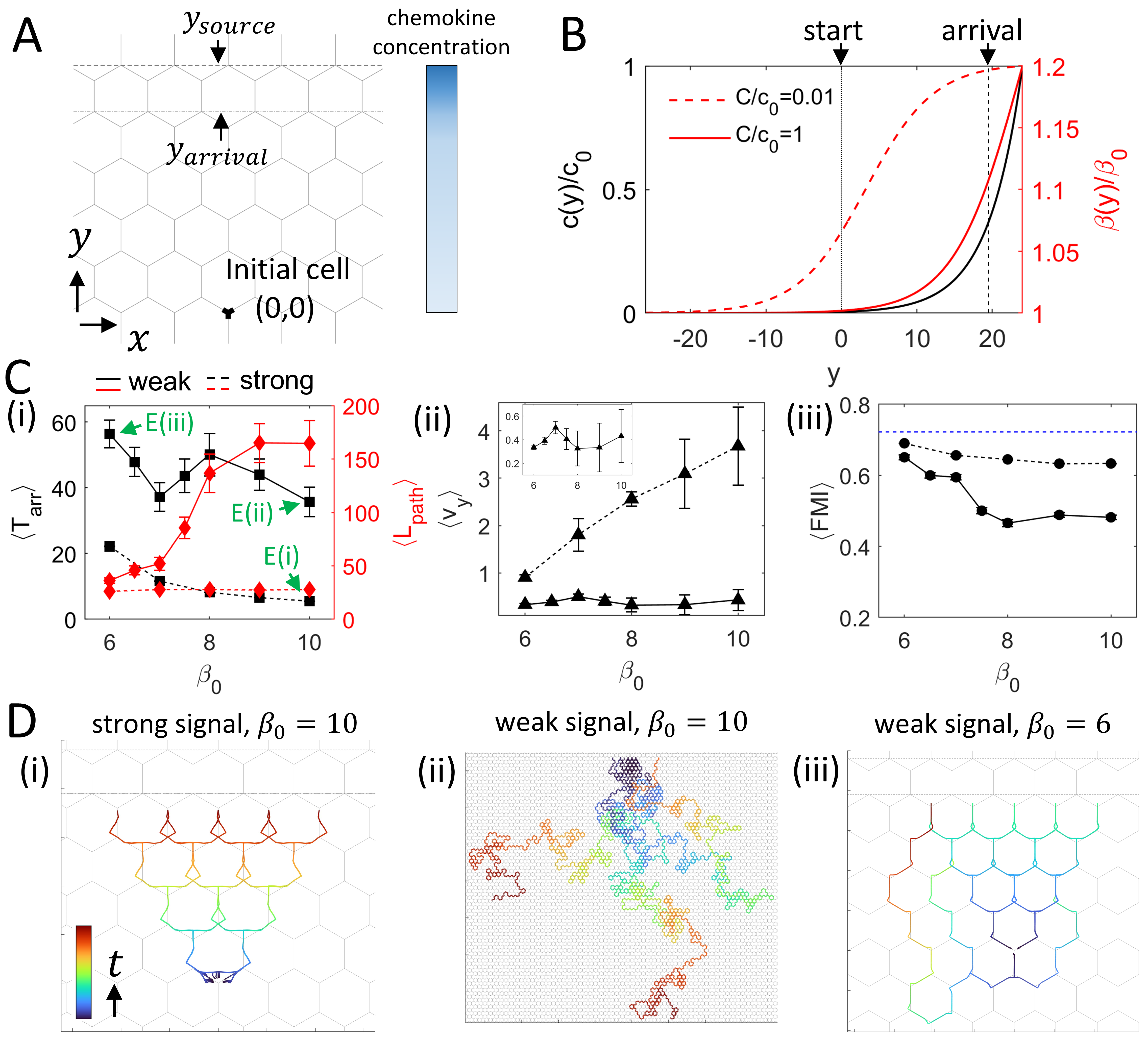}
    \caption{Multiple-junction model with a chemokine line source. (A) Schematic representation of the model. The gray dashed line indicates the position of the chemokine source, while the gray dash-dotted line marks the arrival position. (B) Normalized chemokine concentration, \(c(y)/c_0\) (black line), and the enhancement of actin polymerization speed, \(\beta(y)/\beta_0\), as functions of \(y\) in the strong signal regime (red dashed line) and the weak signal regime (red solid line). (C) (i) \(\langle T_{arr} \rangle\) and \(\langle L_{path} \rangle\) as functions of \(\beta_0\). (ii) \(\langle v_y \rangle\) as a function of \(\beta_0\). The inset corresponds to the weak signal regime. (iii) Forward Migration Index \(\langle FMI \rangle\) (black solid line) and maximal theoretical \(FMI\) (blue dashed line) as functions of \(\beta_0\). Black solid lines and dashed lines correspond to the weak and strong signal regimes in (B), respectively. (D) Typical trajectories of the cell's C.O.M. during a simulation with (i) strong signal, \(\beta_0=10.0\), (ii) weak signal, \(\beta_0=10.0\), and (iii) weak signal, \(\beta_0=6.0\). Maximal simulation time: \(T=1000\). Other key parameters: \(\epsilon=0.2\), \(d=3\), \(\sigma=0.1\).}
    \label{fig3}
\end{figure}

In this study, we model a chemokine source along a one-dimensional line located at \(y_{source} = 8d\), where $d$ is the edge length of the hexagons in the network. The cell's arrival is defined as the point where any of its arm tips reaches \(y_{arr} = 6.5d\) (black dash-dotted line in the sketch in Fig.~\ref{fig3}A). The chemokine concentration is therefore written as:

\begin{equation}
c(y) = c_0 \, \exp\left(-\frac{|y-y_{\text{source}}|}{y_0}\right)
\label{cy}
\end{equation}
\noindent
where \(y\) is the \(y\)-coordinate of the arm tip.

We fixed the chemokine profile parameters (\(\epsilon = 0.2\) and \(y_0 = 1.5d\)) and vary the parameter \(C/c_0\) which represents different saturation regimes of the cells in the chemokine gradient. For \(C/c_0 = 0.01\), the chemokine source attracts the cell even when it is far away (Fig.~\ref{fig3}B), representing the strong signal (high saturation) regime. For \(C/c_0 = 1\), the chemokine gradient is only effective when the cell is very close to the source (Fig.~\ref{fig3}B), representing the weak signal (low saturation) regime. At the start of each simulation, the cell is symmetrically positioned at the origin \((x, y) = (0, 0)\) and allowed to spread, polarize and migrate. The simulation ends when the cell reaches the source, or when the maximal simulation time is reached \(T_{max} = 1000\). 

We evaluated the efficiency of the chemotactic migration by plotting the mean arrival time, \(T_{arr}\), which reflects both the speed and accuracy of the DDM during the migration (Fig.~\ref{fig3}C(i)). Additionally, we calculated the mean distance of the cell's center-of-mass (C.O.M.) trajectory, \(L_{\text{path}}\) (Fig.~\ref{fig3}C(i), mean speed towards the chemokine source (\(\langle v_y \rangle\), Fig.~\ref{fig3}C(ii)) and forward migration index (FMI, Fig.~\ref{fig3}C(iii)).

In the strong signal regime, the cells rarely take wrong turns and migrate along directed paths to the chemokine source (see typical trajectories for $\beta_0=10$ in Fig.~\ref{fig3}D(i)). As a result, \(\langle L_{path} \rangle\) is small and independent of $\beta_0$ (Fig.~\ref{fig3}C(i), red line). The high accuracy is also reflected in \(\langle v_y \rangle\) (Fig.~\ref{fig3}C(ii)), which is increasing almost linearly with \(\beta_0\). Similarly, \(\langle T_{arr} \rangle\) decreases monotonically as \(\beta_0\) increases (Fig.~\ref{fig3}C(i), black dashed line), since the cell's speed increases with \(\beta_0\), without reducing its accuracy. 

In the weak signal regime, we find a more complex behavior as a function of increasing $\beta_0$. At the lowest values of $\beta_0$, we find that \(\langle L_{path} \rangle\) and FMI are very similar although \(\langle T_{arr} \rangle\) is longer for strong signals (Fig.~\ref{fig3}), indicating that these slowest cells are very accurate in their chemotaxis migration even in response to a very weak signal.

As \(\beta_0\) increases, the \(\langle L_{path} \rangle\) increases monotonically (Fig.~\ref{fig3}C(i), red solid line), indicating a higher probability of making incorrect DDM, as turning away from the chemokine source. This fits with the speed-accuracy relation we found on a single junction in Fig.~\ref{fig2}B. It is most clearly shown by plotting typical trajectories of the cell's C.O.M. at large $\beta_0=10$ (Fig.~~\ref{fig3}D(ii)). 

At intermediate values, around \(\beta_0\), we find a significant minimum of \(\langle T_{arr} \rangle\) (Fig.~\ref{fig3}C(i), black solid line). This is due to the slower increase in the error rate (quantified by the FMI in Fig.~\ref{fig3}C(iii)) in comparison to the increase in speed, with an increase of \(\beta_0\) (Fig.~\ref{fig3}C(ii)). At higher $\beta_0$ values, the error rate increases faster than the speed, which increases \(\langle T_{arr} \rangle\) that reaches a peak around $\beta_0\sim8$ (Fig.~\ref{fig3}C(i), black solid line). The non-monotonous behavior of \(\langle T_{arr} \rangle\) that we find, largely follows the speed-accuracy trends that we found on a single junction (Fig.~\ref{fig2}B(i)).

However, despite the wrong turns, the higher overall speed for the high values of \(\beta_0\sim10\) results in a low \(\langle T_{arr} \rangle\) (Fig.~\ref{fig3}C(i), black solid line), which is also reflected by \(\langle v_y \rangle\) (Fig.~\ref{fig3}C(ii), solid line). Note that for the very small fraction of cells ($\sim1-2\%$) that do not arrive at the target within the \(T_{\text{max}}\), we assigned \(T_{\text{arr}}= 1000\).

In the SI section S-4, we investigated the effect of cellular internal noise on chemotactic migration in both the strong and weak signal regimes (Fig.~S-4A,B). We also investigated the migration dynamics of cells on large grids (\(d = 7.5\)) (Fig.~S-4C,D). In this case, we found that cells with large \(\beta_0\) values occasionally exhibit slow-mode events, similar to the behavior observed in the one-junction case (Fig.~\ref{fig2}C(iii)). We show that the slow-mode appears along the trajectory, which causes the mean arrival time to increase and the FMI (accuracy) to improve in the strong-signal regime (see Fig.~S-4D,F). Note that a similar change in the migration dynamics can be induced by keeping the grid size fixed while the cell length is modified, for example by changing the cell's contractility and stiffness parameter $k$. The slow-mode events are also shown in simulations towards a point-like chemokine source (Figs.S-9,\ref{fig5}), and have two types: either the two long arms form in the direction of migration (Fig.\ref{fig5}) as in a single junction (Fig.~\ref{fig2}C(iii)), or one of the arms form along the direction from which the cell arrived at the junction, following a stick-slip event (Fig.S-9).
%In Fig.~S-4F, under the large-grid condition that induces slow-mode events in cells with high-activity \(\beta_0=9.5\), we investigated the effect of grid size on chemotactic migration in the strong signal regime. In Fig.~S-3G, we show typical trajectories of the cell's C.O.M. during migration in which cell exhibits several slow events.
We also give additional detailed analysis of the distribution of arrival times, and how they are related to the errors that the cell performs along its migration path (SI section S-5, Fig.S-5).

From the theoretical studies above, we can summarize the following conclusions regarding the best strategies that an immune cell may adopt in order to arrive to a site of a wound or infection using chemotaxis:
\begin{itemize}
    \item In the presence of a strong signal, as may be expected close to the target site (wound or infection), cells with higher actin polymerization activity $\beta_0$ arrive faster, and therefore can prevent bacterial entry most effectively (Fig.~\ref{fig3}C(i)).
    \item In the presence of a weak signal, far from the target site, we find that the slowest cells are able to maintain accurate and directed chemotaxis migration paths. Despite the high accuracy, the arrival time for these slow-moving cells is relatively long.
    \item When the signal is weak, the fastest arrival times are found for a narrow range of intermediate cellular activities ($\beta_0\sim7$) (Fig.~\ref{fig3}D(i)). In this intermediate regime, the chemotaxis is most efficient. This is demonstrated by short arrival times and low energy expenditure (which may allow them to have a longer lifespan or remain more potent, for example).
    \item For weak signals far from the target site, cells with high $\beta_0$ exhibit a high percentage of meandering paths due to lower chemotactic accuracy (compared to slower cells). Nevertheless, they can have a relatively short mean arrival time, due to their high speeds. On grids of larger size these highly active cells may get further slowed down due to getting transiently trapped in slow-mode events (Figs.~S-3,S-8,\ref{fig5}). Efficient recruitment of these high-activity cells from large distances may therefore not rely only on the initial signal from the wound/infection site, but utilize further strategies to enhance the chemical guidance signal.
\end{itemize}

\section*{Comparison of the model with the chemotaxis of neutrophils}

\begin{figure}[h!]
    \centering
    \includegraphics[width=1\textwidth]{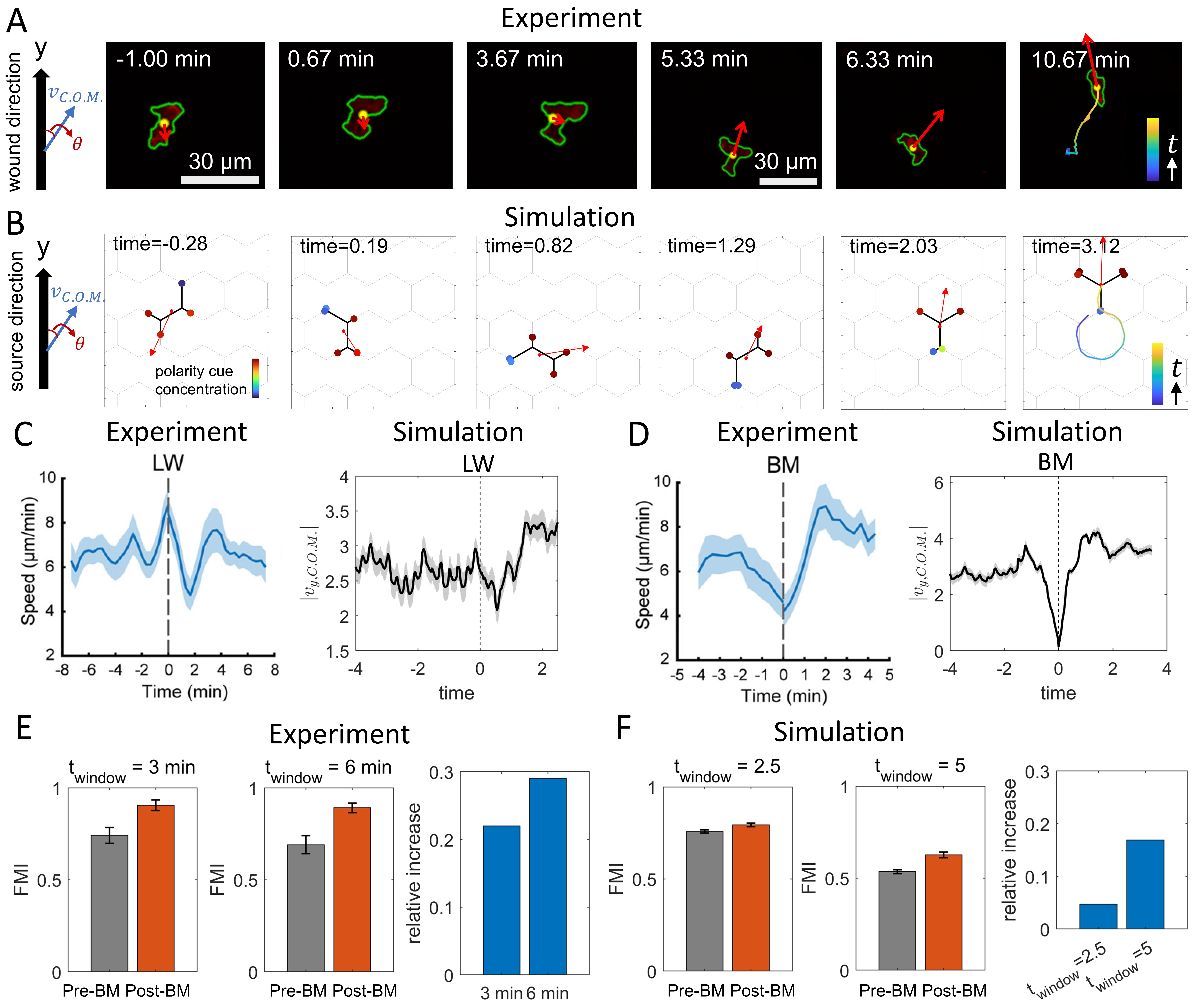}
    \caption{Comparisons between the migration dynamics in simulations and neutrophils in zebra-fish larva \cite{georgantzoglou2022two}.(A-B) Snapshots of experimental and simulated cell migration processes. We pick examples where the cell was moving roughly away from the signal, which is located at the positive $y$-direction (large arrow in leftmost panels). At time $t=0$ the laser-induced wound is applied in the experiments, and the gradient in chemokine is introduced in the simulations. At the final time panel we show the trajectory of the cell's center of mass (C.O.M.). Red arrows show the instantaneous velocity vector. (C) Left: Experimental cell's C.O.M. velocity before and after the LW event. Data is obtained from 21 cells. Right: Simulated cell's center-of-mass velocity in the \(y\)-direction, \(|v_{y,\text{C.O.M.}}|\), before and after the LW event. Data is obtained from 60 simulation runs. (D) Left: Experimental cell's center-of-mass velocity before and after the time of the beginning of movement (BM). Data is obtained from 18 cells. Right: Simulated cell's center-of-mass velocity in the \(y\)-direction, \(|v_{y,\text{C.O.M.}}|\), before and after the BM time. Data is obtained from 60 simulation runs. (E) FMI of the experimental cell's center of mass (C.O.M.) within a specific time window (left two panels) and the relative increase after the BM time compared to before the BM time. (F) FMI of the simulated cell's C.O.M. within a specific time window (left two panels) and the relative increase after the BM time compared to before the BM time (rightmost panel). Key parameters: \(\epsilon=0.1, d=3, \beta_0=12, \sigma=0.5\).}
    \label{fig4}
\end{figure}

Next, we compare our theoretical model with \textit{in-vivo} experimental data of neutrophils migrating towards laser-inflicted wounds (LW) \cite{poplimont2020neutrophil} in the fin of a zebrafish larva \cite{georgantzoglou2022two}.

To simplify the analysis of the chemotaxis within our theoretical model, we assume a spatially linear variation of the chemokine concentration (Fig.S-6A), increasing along the \(y\)-axis (for more details of the choice of parameters for the concentration profile see SI section S-6). 

In Fig.~\ref{fig4}A, we show an example of the dynamics of a single migrating neutrophil, which initially moves away from the LW (the LW is located at the $+y$-direction). The cell is observed to respond to the chemical signal from the LW by slowing down, pausing, and rotating towards the LW. The dynamic of the cell's speed along the $y$-direction, and its angle with respect to the $y$-direction, during this process are shown in Fig.S-5C. While migrating towards the LW, the neutrophil displays large speeds and large directional variability (see time$>6$ min in Fig.S-6C). Further properties of the simulated cell's response to the chemokine gradient, are discussed in SI section S-6.

In Fig.~\ref{fig4}C,D we compare the dynamics of the cells in experiments and simulations, when averaged over many trajectories (21 and 18 cells for Fig.~\ref{fig4}C and D in the experiments \cite{georgantzoglou2022two}), and 60 simulations). In Fig.~\ref{fig4}C, the average cell speed in experiments and simulations are plotted with respect to the LW time. In the simlation, the speed towards the chemokine source in the \(y\)-direction is used to eliminate the oscillations in the perpendicular direction (\(x\)-direction) caused by the hexagonal grid topology, which induces significant fluctuations in the overall velocity. In Fig.~\ref{fig4}D, the overall experimental speed and the \(y\)-direction simulation speed are plotted with respect to the time where the cell was first observed to start moving towards the source (denoted as "beginning of movement" BM time, see \cite{georgantzoglou2022two} and SI Section S-6). In Fig.~\ref{fig4}C,D the transient slowing down after the chemokine gradient is introduced is clearly observed, in both experiments and simulations. Overall, the simulations are found to reproduce the qualitative features of how neutrophils respond to the chemokine source, and provide a mechanistic explanation for the origin of these dynamics.

%\section*{Comparison of the model with the chemotaxis properties of neutrophils in-vivo}

Let us next note the mapping of length and time scales between the experiments and model. In the experiments the cells were $\sim$ $20$ $\mu m$ in length (Fig.~\ref{fig4}A) and moved at speed of $\sim$ $6$ $\mu m$/min on average (Fig.~\ref{fig4}C). The simulated cell (for $\beta_0=12$) has an average length $9.38$ and speed of $3.59$, both in simulation units. This means that the experimental cell covers about 1 cell length in a $3$ min interval. This corresponds to choosing a simulation time interval of $\sim2.5$, such that the simulated cell also covers a distance which corresponds to about 1 cell length. Furthermore, to facilitate the comparison between the theoretical model and the experiments, we present all the data as dimensionless or normalized quantities.

In Fig.~\ref{fig4}E,F, we compare the effects of the chemokine presence on the FMI of the neutrophils \cite{georgantzoglou2022two} with our simulation model. We use this experimental data to calibrate the value of the parameter $\beta_0$ in our simulations. The FMI of the wild-type cells is observed to significantly increase after the BM time (by $\sim20\%-30\%$, Fig.~\ref{fig4}E). This behavior is captured by the model only when we are in the regime of large $\beta_0$.

We next compared the model predictions to the observations of chemotaxis on cells treated with various drugs that inhibit the cytoskeletal activity. We incorporate the inhibition of actin polymerization or myosin-II activity in our model through changes to the model parameters, as was done previously \cite{ron2024emergent}. Comparing the model to the experiments on these drug-treated cells (Fig.~S-7) gives good qualitative agreement, and points to the WT cells residing in the high-$\beta_0$ regime of our model. 

\begin{figure}[h!]
    \centering
    \includegraphics[width=1\textwidth]{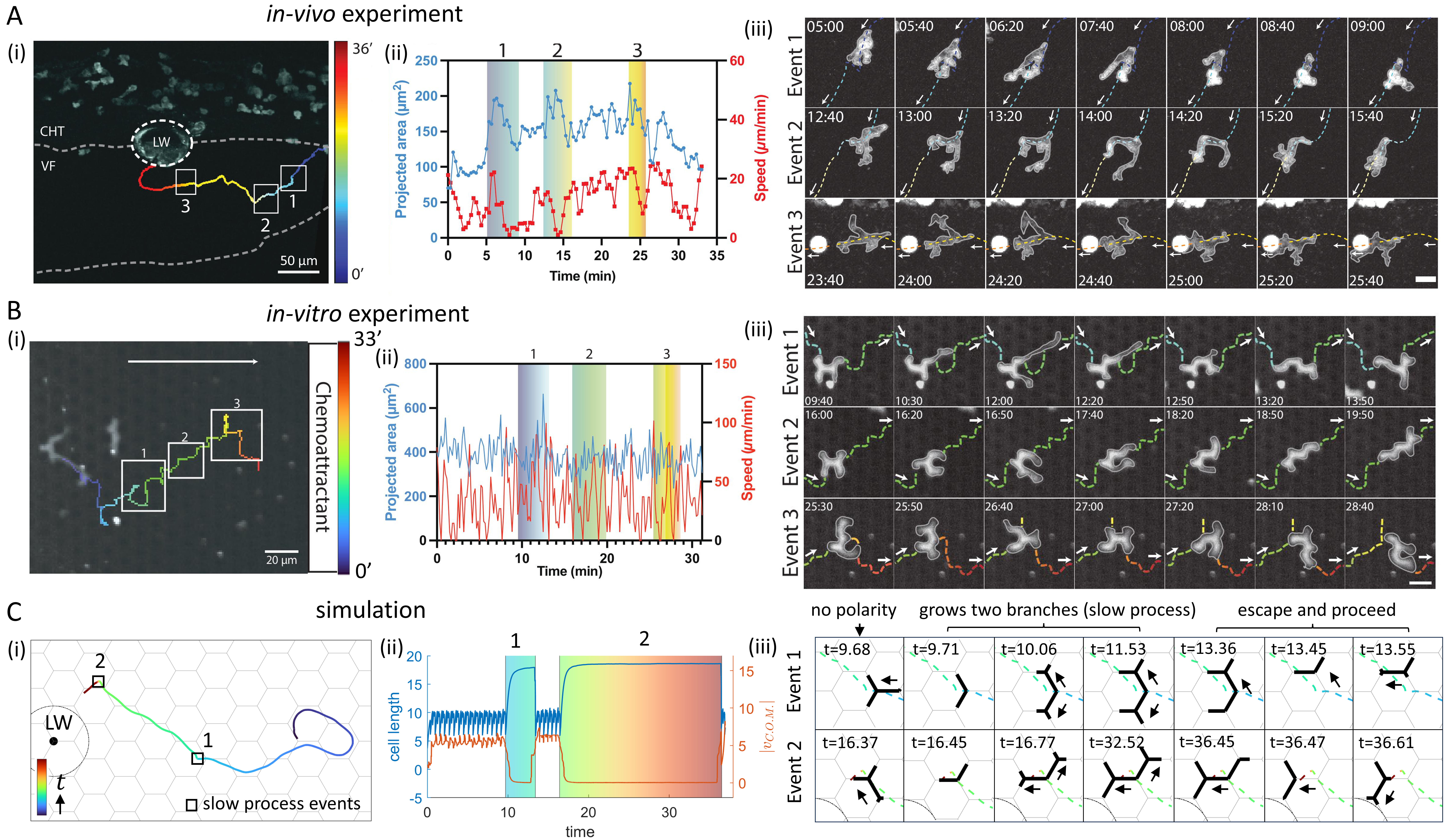}
    \caption{Comparisons between the migration dynamics in simulations and neutrophils. (A) (i) A representative trajectory of a neutrophil migrating toward the LW, with the trajectory color indicating migration time. White boxes highlight time periods along the trajectory in which slow-mode events occur, corresponding to the images in (iii). (ii) Dynamics of the projected area and the overall speed of the neutrophil during migration. The three colored regions correspond to the three slow-mode events marked in (i). (iii) Snapshots of the neutrophil during the three slow-mode events. (B) Same panels as in (A) for a PLB-985 (promyelocytic leukemia blasts)  cell migrating \textit{in-vitro} up a linear chemokine gradient, while confined within a regular lattice of pillars. (C) (i) A representative simulation trajectory of a cell migrating toward the LW, with the trajectory color indicating migration time. Black boxes highlight time periods along the trajectory in which slow-mode events occur, corresponding to the color regions in (ii) and the images in (iii). (ii) Dynamics of cell length and its C.O.M. speed during migration. The two colored regions correspond to the three slow-mode events marked in (i). (F) Simulation snapshots of the cell during the two slow-mode events (marked by boxes in (i)). Parameters: \(C/c_0=0.01, \epsilon=0.1, d=3.7, \beta_0=12.6, \sigma=0.6\).}
    \label{fig5}
\end{figure}

Our conclusion from the comparisons above, that the neutrophils correspond to the high-activity cells of our model, is further supported by the following features of their migration. When the neutrophils start their trajectory further away from the LW, we often find that cells exhibit highly meandering paths (Fig.~\ref{fig5}A(i)). Such meandering paths are found in our model only for the high-$\beta_0$ regime (see Fig.~\ref{fig3}D). Furthermore, we can often identify that the cells display the slow-mode behavior along these trajectories (Fig.~\ref{fig2}C(iii)), as shown in Fig.~\ref{fig5}A(ii,iii) (the prevalence of these events is quantified in Fig.~S-8). Very similar features of cell migration and shape dynamics are observed for a PLB-985 (promyelocytic leukemia blasts) cell migrating \textit{in-vitro} towards a linear chemokine gradient (Fig.~\ref{fig5}B), while confined within a regular hexagonal lattice of rectangular pillars (see SI section S-2 for more details about these experiments \cite{renkawitz2018micro}). This demonstrates that the migration characteristics are not dependent on specific \textit{in-vivo} interactions of the neutrophils with the surrounding tissue cells, but are general features of the chemotactic migration of fast-moving branched cells within a complex environment. A simulation of a high-activity cell migrating towards the LW predicts meandering paths with slow-mode events (Fig.~\ref{fig5}C). Note that the total cell length increases in the slow-mode events (both in experiments and simulations, Fig.~\ref{fig5}A-C panels (ii)) as the two arms extend, before one of them retracts and allows the cell to resume its migration. Note that in the simulations the slow-mode events tend to be more persistent as the cell is closer to the chemokine source, due to the increased actin activity (Figs.~\ref{fig5}C(ii),S-8).

\section{Conclusion}

We presented here a theoretical model of the chemotaxis of branched cells on networks. The model exhibits a speed-accuracy trade-off, and predicts that fast cells are less accurate than slow cells. The model provides a mechanistic description of how motile cells respond, reorient and migrate towards a chemoattractant source, in good agreement with experimental observations of immune cells \textit{in-vivo}.

Comparing the model to experimental observations of neutrophils migrating \textit{in-vivo} to the site of a laser-inflicted wound in a zebrafish larva fin, and to PLB-985 (promyelocytic leukemia blasts)  cells migrating \textit{in-vitro} within a regular lattice of pillars, suggests that these immune cells correspond to the fast-cell limit of the theoretical model. Cells in this limit have the advantage of minimizing their arrival time to the wound site, if they start close to the wound and receive a strong chemokine signal. This can therefore be a good strategy if the neutrophils are uniformly spread at sufficiently high concentration in the skin, so that there are always some cells close to any wound site. However, such fast moving cells perform less accurate chemotaxis when further from the wound, where the original wound-secreted signal is weak. For efficient recruitment of these far-field neutrophils the immune cells need to utilize different mechanisms, such as secretion of their own chemokine signals. Indeed, neutrophils have evolved such recruitment mechanisms, which have been discovered and are being studied \cite{afonso2012ltb4,reategui2017microscale,song2023molecular,strickland2024self}.

We demonstrated here that our simplified theoretical framework gives a description of both the microscopic dynamics of cell polarization and shape changes, as well as the large-scale migration patterns. This can be a useful tool for deciphering the behavior of motile cells, such as immune or cancer cells, performing chemotaxis inside the complex geometries of living tissues.

\section{Acknowledgements}
N.S.G. is the incumbent of the Lee and William Abramowitz Professorial Chair of Biophysics (Weizmann Institute), and acknowledges support from the Royal Society Wolfson Visiting Fellowship, and Human Frontier Science Program grant RGP0032/2022. Work by M.S., I.W., G.R. and A.G. was supported by the Leverhulme Trust (grant RPG-2021-226) and the European Research Council (ERC) under the Horizon 2020 program and UKRI, Grant agreement No. EP/Y02799X/1. M.S. and I.d.V acknowledge support by the European Research Council (grant ERC-SyG 101071793 to M.S), and thank Jack Merrin for support with microfluidic engineering.

%\clearpage
\bibliography{bibliography}

\providecommand{\noopsort}[1]{}\providecommand{\singleletter}[1]{#1}%
\begin{thebibliography}{22}
\expandafter\ifx\csname natexlab\endcsname\relax\def\natexlab#1{#1}\fi
\expandafter\ifx\csname bibnamefont\endcsname\relax
  \def\bibnamefont#1{#1}\fi
\expandafter\ifx\csname bibfnamefont\endcsname\relax
  \def\bibfnamefont#1{#1}\fi
\expandafter\ifx\csname citenamefont\endcsname\relax
  \def\citenamefont#1{#1}\fi
\expandafter\ifx\csname url\endcsname\relax
  \def\url#1{\texttt{#1}}\fi
\expandafter\ifx\csname urlprefix\endcsname\relax\def\urlprefix{URL }\fi
\providecommand{\bibinfo}[2]{#2}
\providecommand{\eprint}[2][]{\url{#2}}

\bibitem[{\citenamefont{Bodor et~al.}(2020)\citenamefont{Bodor, P{\"o}nisch,
  Endres, and Paluch}}]{bodor2020cell}
\bibinfo{author}{\bibfnamefont{D.~L.} \bibnamefont{Bodor}},
  \bibinfo{author}{\bibfnamefont{W.}~\bibnamefont{P{\"o}nisch}},
  \bibinfo{author}{\bibfnamefont{R.~G.} \bibnamefont{Endres}},
  \bibnamefont{and} \bibinfo{author}{\bibfnamefont{E.~K.}
  \bibnamefont{Paluch}}, \bibinfo{journal}{Developmental cell}
  \textbf{\bibinfo{volume}{52}}, \bibinfo{pages}{550} (\bibinfo{year}{2020}).

\bibitem[{\citenamefont{S{\'a}ez et~al.}(2018)\citenamefont{S{\'a}ez, Barbier,
  Attia, Thiam, Piel, and Vargas}}]{saez2018leukocyte}
\bibinfo{author}{\bibfnamefont{P.~J.} \bibnamefont{S{\'a}ez}},
  \bibinfo{author}{\bibfnamefont{L.}~\bibnamefont{Barbier}},
  \bibinfo{author}{\bibfnamefont{R.}~\bibnamefont{Attia}},
  \bibinfo{author}{\bibfnamefont{H.-R.} \bibnamefont{Thiam}},
  \bibinfo{author}{\bibfnamefont{M.}~\bibnamefont{Piel}}, \bibnamefont{and}
  \bibinfo{author}{\bibfnamefont{P.}~\bibnamefont{Vargas}},
  \bibinfo{journal}{Cell Migration: Methods and Protocols} pp.
  \bibinfo{pages}{361--373} (\bibinfo{year}{2018}).

\bibitem[{\citenamefont{Vargas et~al.}(2017)\citenamefont{Vargas, Barbier,
  S{\'a}ez, and Piel}}]{vargas2017mechanisms}
\bibinfo{author}{\bibfnamefont{P.}~\bibnamefont{Vargas}},
  \bibinfo{author}{\bibfnamefont{L.}~\bibnamefont{Barbier}},
  \bibinfo{author}{\bibfnamefont{P.~J.} \bibnamefont{S{\'a}ez}},
  \bibnamefont{and} \bibinfo{author}{\bibfnamefont{M.}~\bibnamefont{Piel}},
  \bibinfo{journal}{Current Opinion in Cell Biology}
  \textbf{\bibinfo{volume}{48}}, \bibinfo{pages}{72} (\bibinfo{year}{2017}).

\bibitem[{\citenamefont{Georgantzoglou
  et~al.}(2022)\citenamefont{Georgantzoglou, Poplimont, Walker, L{\"a}mmermann,
  and Sarris}}]{georgantzoglou2022two}
\bibinfo{author}{\bibfnamefont{A.}~\bibnamefont{Georgantzoglou}},
  \bibinfo{author}{\bibfnamefont{H.}~\bibnamefont{Poplimont}},
  \bibinfo{author}{\bibfnamefont{H.~A.} \bibnamefont{Walker}},
  \bibinfo{author}{\bibfnamefont{T.}~\bibnamefont{L{\"a}mmermann}},
  \bibnamefont{and} \bibinfo{author}{\bibfnamefont{M.}~\bibnamefont{Sarris}},
  \bibinfo{journal}{Journal of Cell Biology} \textbf{\bibinfo{volume}{221}},
  \bibinfo{pages}{e202103207} (\bibinfo{year}{2022}).

\bibitem[{\citenamefont{Friedl and Alexander}(2011)}]{friedl2011cancer}
\bibinfo{author}{\bibfnamefont{P.}~\bibnamefont{Friedl}} \bibnamefont{and}
  \bibinfo{author}{\bibfnamefont{S.}~\bibnamefont{Alexander}},
  \bibinfo{journal}{Cell} \textbf{\bibinfo{volume}{147}}, \bibinfo{pages}{992}
  (\bibinfo{year}{2011}).

\bibitem[{\citenamefont{Garcia-Arcos et~al.}(2019)\citenamefont{Garcia-Arcos,
  Chabrier, Deygas, Nader, Barbier, S{\'a}ez, Mathur, Vargas, and
  Piel}}]{garcia2019reconstitution}
\bibinfo{author}{\bibfnamefont{J.~M.} \bibnamefont{Garcia-Arcos}},
  \bibinfo{author}{\bibfnamefont{R.}~\bibnamefont{Chabrier}},
  \bibinfo{author}{\bibfnamefont{M.}~\bibnamefont{Deygas}},
  \bibinfo{author}{\bibfnamefont{G.}~\bibnamefont{Nader}},
  \bibinfo{author}{\bibfnamefont{L.}~\bibnamefont{Barbier}},
  \bibinfo{author}{\bibfnamefont{P.~J.} \bibnamefont{S{\'a}ez}},
  \bibinfo{author}{\bibfnamefont{A.}~\bibnamefont{Mathur}},
  \bibinfo{author}{\bibfnamefont{P.}~\bibnamefont{Vargas}}, \bibnamefont{and}
  \bibinfo{author}{\bibfnamefont{M.}~\bibnamefont{Piel}},
  \bibinfo{journal}{Journal of cell science} \textbf{\bibinfo{volume}{132}},
  \bibinfo{pages}{jcs225565} (\bibinfo{year}{2019}).

\bibitem[{\citenamefont{Sarris and Sixt}(2015)}]{sarris2015navigating}
\bibinfo{author}{\bibfnamefont{M.}~\bibnamefont{Sarris}} \bibnamefont{and}
  \bibinfo{author}{\bibfnamefont{M.}~\bibnamefont{Sixt}},
  \bibinfo{journal}{Current opinion in cell biology}
  \textbf{\bibinfo{volume}{36}}, \bibinfo{pages}{93} (\bibinfo{year}{2015}).

\bibitem[{\citenamefont{Yamada and Sixt}(2019)}]{yamada2019mechanisms}
\bibinfo{author}{\bibfnamefont{K.~M.} \bibnamefont{Yamada}} \bibnamefont{and}
  \bibinfo{author}{\bibfnamefont{M.}~\bibnamefont{Sixt}},
  \bibinfo{journal}{Nature Reviews molecular cell biology}
  \textbf{\bibinfo{volume}{20}}, \bibinfo{pages}{738} (\bibinfo{year}{2019}).

\bibitem[{\citenamefont{Renkawitz et~al.}(2018)\citenamefont{Renkawitz,
  Reversat, Leithner, Merrin, and Sixt}}]{renkawitz2018micro}
\bibinfo{author}{\bibfnamefont{J.}~\bibnamefont{Renkawitz}},
  \bibinfo{author}{\bibfnamefont{A.}~\bibnamefont{Reversat}},
  \bibinfo{author}{\bibfnamefont{A.}~\bibnamefont{Leithner}},
  \bibinfo{author}{\bibfnamefont{J.}~\bibnamefont{Merrin}}, \bibnamefont{and}
  \bibinfo{author}{\bibfnamefont{M.}~\bibnamefont{Sixt}}, in
  \emph{\bibinfo{booktitle}{Methods in cell biology}}
  (\bibinfo{publisher}{Elsevier}, \bibinfo{year}{2018}), vol.
  \bibinfo{volume}{147}, pp. \bibinfo{pages}{79--91}.

\bibitem[{\citenamefont{Ron et~al.}(2024)\citenamefont{Ron, Crestani, Kux, Liu,
  Al-Dam, Monzo, Gauthier, S{\'a}ez, and Gov}}]{ron2024emergent}
\bibinfo{author}{\bibfnamefont{J.~E.} \bibnamefont{Ron}},
  \bibinfo{author}{\bibfnamefont{M.}~\bibnamefont{Crestani}},
  \bibinfo{author}{\bibfnamefont{J.~M.} \bibnamefont{Kux}},
  \bibinfo{author}{\bibfnamefont{J.}~\bibnamefont{Liu}},
  \bibinfo{author}{\bibfnamefont{N.}~\bibnamefont{Al-Dam}},
  \bibinfo{author}{\bibfnamefont{P.}~\bibnamefont{Monzo}},
  \bibinfo{author}{\bibfnamefont{N.~C.} \bibnamefont{Gauthier}},
  \bibinfo{author}{\bibfnamefont{P.~J.} \bibnamefont{S{\'a}ez}},
  \bibnamefont{and} \bibinfo{author}{\bibfnamefont{N.~S.} \bibnamefont{Gov}},
  \bibinfo{journal}{Nature Physics} pp. \bibinfo{pages}{1--11}
  (\bibinfo{year}{2024}).

\bibitem[{\citenamefont{Liu et~al.}(2024)\citenamefont{Liu, Boix-Campos, Ron,
  Kux, Gov, and S{\'a}ez}}]{liu2024shape}
\bibinfo{author}{\bibfnamefont{J.}~\bibnamefont{Liu}},
  \bibinfo{author}{\bibfnamefont{J.}~\bibnamefont{Boix-Campos}},
  \bibinfo{author}{\bibfnamefont{J.~E.} \bibnamefont{Ron}},
  \bibinfo{author}{\bibfnamefont{J.~M.} \bibnamefont{Kux}},
  \bibinfo{author}{\bibfnamefont{N.~S.} \bibnamefont{Gov}}, \bibnamefont{and}
  \bibinfo{author}{\bibfnamefont{P.~J.} \bibnamefont{S{\'a}ez}},
  \bibinfo{journal}{arXiv preprint arXiv:2404.00118}  (\bibinfo{year}{2024}).

\bibitem[{\citenamefont{Poplimont et~al.}(2020)\citenamefont{Poplimont,
  Georgantzoglou, Boulch, Walker, Coombs, Papaleonidopoulou, and
  Sarris}}]{poplimont2020neutrophil}
\bibinfo{author}{\bibfnamefont{H.}~\bibnamefont{Poplimont}},
  \bibinfo{author}{\bibfnamefont{A.}~\bibnamefont{Georgantzoglou}},
  \bibinfo{author}{\bibfnamefont{M.}~\bibnamefont{Boulch}},
  \bibinfo{author}{\bibfnamefont{H.~A.} \bibnamefont{Walker}},
  \bibinfo{author}{\bibfnamefont{C.}~\bibnamefont{Coombs}},
  \bibinfo{author}{\bibfnamefont{F.}~\bibnamefont{Papaleonidopoulou}},
  \bibnamefont{and} \bibinfo{author}{\bibfnamefont{M.}~\bibnamefont{Sarris}},
  \bibinfo{journal}{Current Biology} \textbf{\bibinfo{volume}{30}},
  \bibinfo{pages}{2761} (\bibinfo{year}{2020}).

\bibitem[{\citenamefont{Ron et~al.}(2020)\citenamefont{Ron, Monzo, Gauthier,
  Voituriez, and Gov}}]{ron2020one}
\bibinfo{author}{\bibfnamefont{J.~E.} \bibnamefont{Ron}},
  \bibinfo{author}{\bibfnamefont{P.}~\bibnamefont{Monzo}},
  \bibinfo{author}{\bibfnamefont{N.~C.} \bibnamefont{Gauthier}},
  \bibinfo{author}{\bibfnamefont{R.}~\bibnamefont{Voituriez}},
  \bibnamefont{and} \bibinfo{author}{\bibfnamefont{N.~S.} \bibnamefont{Gov}},
  \bibinfo{journal}{Physical Review Research} \textbf{\bibinfo{volume}{2}},
  \bibinfo{pages}{033237} (\bibinfo{year}{2020}).

\bibitem[{\citenamefont{Insall and Machesky}(2009)}]{insall2009actin}
\bibinfo{author}{\bibfnamefont{R.~H.} \bibnamefont{Insall}} \bibnamefont{and}
  \bibinfo{author}{\bibfnamefont{L.~M.} \bibnamefont{Machesky}},
  \bibinfo{journal}{Developmental cell} \textbf{\bibinfo{volume}{17}},
  \bibinfo{pages}{310} (\bibinfo{year}{2009}).

\bibitem[{\citenamefont{Ivaska}(2012)}]{ivaska2012unanchoring}
\bibinfo{author}{\bibfnamefont{J.}~\bibnamefont{Ivaska}},
  \bibinfo{journal}{Nature cell biology} \textbf{\bibinfo{volume}{14}},
  \bibinfo{pages}{981} (\bibinfo{year}{2012}).

\bibitem[{\citenamefont{Ren et~al.}(2000)\citenamefont{Ren, Kiosses, Sieg,
  Otey, Schlaepfer, and Schwartz}}]{ren2000focal}
\bibinfo{author}{\bibfnamefont{X.-D.} \bibnamefont{Ren}},
  \bibinfo{author}{\bibfnamefont{W.~B.} \bibnamefont{Kiosses}},
  \bibinfo{author}{\bibfnamefont{D.~J.} \bibnamefont{Sieg}},
  \bibinfo{author}{\bibfnamefont{C.~A.} \bibnamefont{Otey}},
  \bibinfo{author}{\bibfnamefont{D.~D.} \bibnamefont{Schlaepfer}},
  \bibnamefont{and} \bibinfo{author}{\bibfnamefont{M.~A.}
  \bibnamefont{Schwartz}}, \bibinfo{journal}{Journal of cell science}
  \textbf{\bibinfo{volume}{113}}, \bibinfo{pages}{3673} (\bibinfo{year}{2000}).

\bibitem[{\citenamefont{Berginski et~al.}(2011)\citenamefont{Berginski,
  Vitriol, Hahn, and Gomez}}]{berginski2011high}
\bibinfo{author}{\bibfnamefont{M.~E.} \bibnamefont{Berginski}},
  \bibinfo{author}{\bibfnamefont{E.~A.} \bibnamefont{Vitriol}},
  \bibinfo{author}{\bibfnamefont{K.~M.} \bibnamefont{Hahn}}, \bibnamefont{and}
  \bibinfo{author}{\bibfnamefont{S.~M.} \bibnamefont{Gomez}},
  \bibinfo{journal}{PloS one} \textbf{\bibinfo{volume}{6}},
  \bibinfo{pages}{e22025} (\bibinfo{year}{2011}).

\bibitem[{\citenamefont{Gupton and
  Waterman-Storer}(2006)}]{gupton2006spatiotemporal}
\bibinfo{author}{\bibfnamefont{S.~L.} \bibnamefont{Gupton}} \bibnamefont{and}
  \bibinfo{author}{\bibfnamefont{C.~M.} \bibnamefont{Waterman-Storer}},
  \bibinfo{journal}{Cell} \textbf{\bibinfo{volume}{125}}, \bibinfo{pages}{1361}
  (\bibinfo{year}{2006}).

\bibitem[{\citenamefont{Afonso et~al.}(2012)\citenamefont{Afonso,
  Janka-Junttila, Lee, McCann, Oliver, Aamer, Losert, Cicerone, and
  Parent}}]{afonso2012ltb4}
\bibinfo{author}{\bibfnamefont{P.~V.} \bibnamefont{Afonso}},
  \bibinfo{author}{\bibfnamefont{M.}~\bibnamefont{Janka-Junttila}},
  \bibinfo{author}{\bibfnamefont{Y.~J.} \bibnamefont{Lee}},
  \bibinfo{author}{\bibfnamefont{C.~P.} \bibnamefont{McCann}},
  \bibinfo{author}{\bibfnamefont{C.~M.} \bibnamefont{Oliver}},
  \bibinfo{author}{\bibfnamefont{K.~A.} \bibnamefont{Aamer}},
  \bibinfo{author}{\bibfnamefont{W.}~\bibnamefont{Losert}},
  \bibinfo{author}{\bibfnamefont{M.~T.} \bibnamefont{Cicerone}},
  \bibnamefont{and} \bibinfo{author}{\bibfnamefont{C.~A.}
  \bibnamefont{Parent}}, \bibinfo{journal}{Developmental cell}
  \textbf{\bibinfo{volume}{22}}, \bibinfo{pages}{1079} (\bibinfo{year}{2012}).

\bibitem[{\citenamefont{Re{\'a}tegui et~al.}(2017)\citenamefont{Re{\'a}tegui,
  Jalali, Khankhel, Wong, Cho, Lee, Serhan, Dalli, Elliott, and
  Irimia}}]{reategui2017microscale}
\bibinfo{author}{\bibfnamefont{E.}~\bibnamefont{Re{\'a}tegui}},
  \bibinfo{author}{\bibfnamefont{F.}~\bibnamefont{Jalali}},
  \bibinfo{author}{\bibfnamefont{A.~H.} \bibnamefont{Khankhel}},
  \bibinfo{author}{\bibfnamefont{E.}~\bibnamefont{Wong}},
  \bibinfo{author}{\bibfnamefont{H.}~\bibnamefont{Cho}},
  \bibinfo{author}{\bibfnamefont{J.}~\bibnamefont{Lee}},
  \bibinfo{author}{\bibfnamefont{C.~N.} \bibnamefont{Serhan}},
  \bibinfo{author}{\bibfnamefont{J.}~\bibnamefont{Dalli}},
  \bibinfo{author}{\bibfnamefont{H.}~\bibnamefont{Elliott}}, \bibnamefont{and}
  \bibinfo{author}{\bibfnamefont{D.}~\bibnamefont{Irimia}},
  \bibinfo{journal}{Nature biomedical engineering}
  \textbf{\bibinfo{volume}{1}}, \bibinfo{pages}{0094} (\bibinfo{year}{2017}).

\bibitem[{\citenamefont{Song et~al.}(2023)\citenamefont{Song, Bhattacharya,
  Clemens, and Dinauer}}]{song2023molecular}
\bibinfo{author}{\bibfnamefont{Z.}~\bibnamefont{Song}},
  \bibinfo{author}{\bibfnamefont{S.}~\bibnamefont{Bhattacharya}},
  \bibinfo{author}{\bibfnamefont{R.~A.} \bibnamefont{Clemens}},
  \bibnamefont{and} \bibinfo{author}{\bibfnamefont{M.~C.}
  \bibnamefont{Dinauer}}, \bibinfo{journal}{Iscience}
  \textbf{\bibinfo{volume}{26}} (\bibinfo{year}{2023}).

\bibitem[{\citenamefont{Strickland et~al.}(2024)\citenamefont{Strickland, Pan,
  Godfrey, Kim, Hopke, Ji, Degrange, Villavicencio, Mansour, Zerbe
  et~al.}}]{strickland2024self}
\bibinfo{author}{\bibfnamefont{E.}~\bibnamefont{Strickland}},
  \bibinfo{author}{\bibfnamefont{D.}~\bibnamefont{Pan}},
  \bibinfo{author}{\bibfnamefont{C.}~\bibnamefont{Godfrey}},
  \bibinfo{author}{\bibfnamefont{J.~S.} \bibnamefont{Kim}},
  \bibinfo{author}{\bibfnamefont{A.}~\bibnamefont{Hopke}},
  \bibinfo{author}{\bibfnamefont{W.}~\bibnamefont{Ji}},
  \bibinfo{author}{\bibfnamefont{M.}~\bibnamefont{Degrange}},
  \bibinfo{author}{\bibfnamefont{B.}~\bibnamefont{Villavicencio}},
  \bibinfo{author}{\bibfnamefont{M.~K.} \bibnamefont{Mansour}},
  \bibinfo{author}{\bibfnamefont{C.~S.} \bibnamefont{Zerbe}},
  \bibnamefont{et~al.}, \bibinfo{journal}{Developmental Cell}
  \textbf{\bibinfo{volume}{59}}, \bibinfo{pages}{2659} (\bibinfo{year}{2024}).

\end{thebibliography}


\providecommand{\noopsort}[1]{}\providecommand{\singleletter}[1]{#1}%
\begin{thebibliography}{10}
\expandafter\ifx\csname natexlab\endcsname\relax\def\natexlab#1{#1}\fi
\expandafter\ifx\csname bibnamefont\endcsname\relax
  \def\bibnamefont#1{#1}\fi
\expandafter\ifx\csname bibfnamefont\endcsname\relax
  \def\bibfnamefont#1{#1}\fi
\expandafter\ifx\csname citenamefont\endcsname\relax
  \def\citenamefont#1{#1}\fi
\expandafter\ifx\csname url\endcsname\relax
  \def\url#1{\texttt{#1}}\fi
\expandafter\ifx\csname urlprefix\endcsname\relax\def\urlprefix{URL }\fi
\providecommand{\bibinfo}[2]{#2}
\providecommand{\eprint}[2][]{\url{#2}}

\bibitem[{\citenamefont{Westerfield}(2007)}]{westerfield2007zebrafish}
\bibinfo{author}{\bibfnamefont{M.}~\bibnamefont{Westerfield}}, \emph{\bibinfo{title}{The Zebrafish Book; A guide for the laboratory use of zebrafish (Danio rerio)}} (\bibinfo{year}{2007}).

\bibitem[{\citenamefont{Williantarra et~al.}(2024)\citenamefont{Williantarra, Georgantzoglou, and Sarris}}]{williantarra2024visualising}
\bibinfo{author}{\bibfnamefont{I.}~\bibnamefont{Williantarra}}, \bibinfo{author}{\bibfnamefont{A.}~\bibnamefont{Georgantzoglou}}, \bibnamefont{and} \bibinfo{author}{\bibfnamefont{M.}~\bibnamefont{Sarris}}, \bibinfo{journal}{Bio-protocol} \textbf{\bibinfo{volume}{14}}, \bibinfo{pages}{e4997} (\bibinfo{year}{2024}).

\bibitem[{\citenamefont{Poplimont et~al.}(2020)\citenamefont{Poplimont, Georgantzoglou, Boulch, Walker, Coombs, Papaleonidopoulou, and Sarris}}]{poplimont2020neutrophil}
\bibinfo{author}{\bibfnamefont{H.}~\bibnamefont{Poplimont}}, \bibinfo{author}{\bibfnamefont{A.}~\bibnamefont{Georgantzoglou}}, \bibinfo{author}{\bibfnamefont{M.}~\bibnamefont{Boulch}}, \bibinfo{author}{\bibfnamefont{H.~A.} \bibnamefont{Walker}}, \bibinfo{author}{\bibfnamefont{C.}~\bibnamefont{Coombs}}, \bibinfo{author}{\bibfnamefont{F.}~\bibnamefont{Papaleonidopoulou}}, \bibnamefont{and} \bibinfo{author}{\bibfnamefont{M.}~\bibnamefont{Sarris}}, \bibinfo{journal}{Current Biology} \textbf{\bibinfo{volume}{30}}, \bibinfo{pages}{2761} (\bibinfo{year}{2020}).

\bibitem[{\citenamefont{Chen et~al.}(2013)\citenamefont{Chen, Wardill, Sun, Pulver, Renninger, Baohan, Schreiter, Kerr, Orger, Jayaraman et~al.}}]{chen2013ultrasensitive}
\bibinfo{author}{\bibfnamefont{T.-W.} \bibnamefont{Chen}}, \bibinfo{author}{\bibfnamefont{T.~J.} \bibnamefont{Wardill}}, \bibinfo{author}{\bibfnamefont{Y.}~\bibnamefont{Sun}}, \bibinfo{author}{\bibfnamefont{S.~R.} \bibnamefont{Pulver}}, \bibinfo{author}{\bibfnamefont{S.~L.} \bibnamefont{Renninger}}, \bibinfo{author}{\bibfnamefont{A.}~\bibnamefont{Baohan}}, \bibinfo{author}{\bibfnamefont{E.~R.} \bibnamefont{Schreiter}}, \bibinfo{author}{\bibfnamefont{R.~A.} \bibnamefont{Kerr}}, \bibinfo{author}{\bibfnamefont{M.~B.} \bibnamefont{Orger}}, \bibinfo{author}{\bibfnamefont{V.}~\bibnamefont{Jayaraman}}, \bibnamefont{et~al.}, \bibinfo{journal}{Nature} \textbf{\bibinfo{volume}{499}}, \bibinfo{pages}{295} (\bibinfo{year}{2013}).

\bibitem[{\citenamefont{Schindelin et~al.}(2012)\citenamefont{Schindelin, Arganda-Carreras, Frise, Kaynig, Longair, Pietzsch, Preibisch, Rueden, Saalfeld, Schmid et~al.}}]{schindelin2012fiji}
\bibinfo{author}{\bibfnamefont{J.}~\bibnamefont{Schindelin}}, \bibinfo{author}{\bibfnamefont{I.}~\bibnamefont{Arganda-Carreras}}, \bibinfo{author}{\bibfnamefont{E.}~\bibnamefont{Frise}}, \bibinfo{author}{\bibfnamefont{V.}~\bibnamefont{Kaynig}}, \bibinfo{author}{\bibfnamefont{M.}~\bibnamefont{Longair}}, \bibinfo{author}{\bibfnamefont{T.}~\bibnamefont{Pietzsch}}, \bibinfo{author}{\bibfnamefont{S.}~\bibnamefont{Preibisch}}, \bibinfo{author}{\bibfnamefont{C.}~\bibnamefont{Rueden}}, \bibinfo{author}{\bibfnamefont{S.}~\bibnamefont{Saalfeld}}, \bibinfo{author}{\bibfnamefont{B.}~\bibnamefont{Schmid}}, \bibnamefont{et~al.}, \bibinfo{journal}{Nature methods} \textbf{\bibinfo{volume}{9}}, \bibinfo{pages}{676} (\bibinfo{year}{2012}).

\bibitem[{\citenamefont{Georgantzoglou et~al.}(2022)\citenamefont{Georgantzoglou, Poplimont, Walker, L{\"a}mmermann, and Sarris}}]{georgantzoglou2022two}
\bibinfo{author}{\bibfnamefont{A.}~\bibnamefont{Georgantzoglou}}, \bibinfo{author}{\bibfnamefont{H.}~\bibnamefont{Poplimont}}, \bibinfo{author}{\bibfnamefont{H.~A.} \bibnamefont{Walker}}, \bibinfo{author}{\bibfnamefont{T.}~\bibnamefont{L{\"a}mmermann}}, \bibnamefont{and} \bibinfo{author}{\bibfnamefont{M.}~\bibnamefont{Sarris}}, \bibinfo{journal}{Journal of Cell Biology} \textbf{\bibinfo{volume}{221}}, \bibinfo{pages}{e202103207} (\bibinfo{year}{2022}).

\bibitem[{\citenamefont{Renkawitz et~al.}(2018)\citenamefont{Renkawitz, Reversat, Leithner, Merrin, and Sixt}}]{renkawitz2018micro}
\bibinfo{author}{\bibfnamefont{J.}~\bibnamefont{Renkawitz}}, \bibinfo{author}{\bibfnamefont{A.}~\bibnamefont{Reversat}}, \bibinfo{author}{\bibfnamefont{A.}~\bibnamefont{Leithner}}, \bibinfo{author}{\bibfnamefont{J.}~\bibnamefont{Merrin}}, \bibnamefont{and} \bibinfo{author}{\bibfnamefont{M.}~\bibnamefont{Sixt}}, in \emph{\bibinfo{booktitle}{Methods in cell biology}} (\bibinfo{publisher}{Elsevier}, \bibinfo{year}{2018}), vol. \bibinfo{volume}{147}, pp. \bibinfo{pages}{79--91}.

\bibitem[{\citenamefont{Liu et~al.}(2024)\citenamefont{Liu, Boix-Campos, Ron, Kux, Gov, and S{\'a}ez}}]{liu2024shape}
\bibinfo{author}{\bibfnamefont{J.}~\bibnamefont{Liu}}, \bibinfo{author}{\bibfnamefont{J.}~\bibnamefont{Boix-Campos}}, \bibinfo{author}{\bibfnamefont{J.~E.} \bibnamefont{Ron}}, \bibinfo{author}{\bibfnamefont{J.~M.} \bibnamefont{Kux}}, \bibinfo{author}{\bibfnamefont{N.~S.} \bibnamefont{Gov}}, \bibnamefont{and} \bibinfo{author}{\bibfnamefont{P.~J.} \bibnamefont{S{\'a}ez}}, \bibinfo{journal}{arXiv preprint arXiv:2404.00118}  (\bibinfo{year}{2024}).

\bibitem[{\citenamefont{Mukherjee et~al.}(2023)\citenamefont{Mukherjee, Ron, Hu, Nishimura, Hanawa-Suetsugu, Behkam, Mimori-Kiyosue, Gov, Suetsugu, and Nain}}]{mukherjee2023actin}
\bibinfo{author}{\bibfnamefont{A.}~\bibnamefont{Mukherjee}}, \bibinfo{author}{\bibfnamefont{J.~E.} \bibnamefont{Ron}}, \bibinfo{author}{\bibfnamefont{H.~T.} \bibnamefont{Hu}}, \bibinfo{author}{\bibfnamefont{T.}~\bibnamefont{Nishimura}}, \bibinfo{author}{\bibfnamefont{K.}~\bibnamefont{Hanawa-Suetsugu}}, \bibinfo{author}{\bibfnamefont{B.}~\bibnamefont{Behkam}}, \bibinfo{author}{\bibfnamefont{Y.}~\bibnamefont{Mimori-Kiyosue}}, \bibinfo{author}{\bibfnamefont{N.~S.} \bibnamefont{Gov}}, \bibinfo{author}{\bibfnamefont{S.}~\bibnamefont{Suetsugu}}, \bibnamefont{and} \bibinfo{author}{\bibfnamefont{A.~S.} \bibnamefont{Nain}}, \bibinfo{journal}{Advanced science} \textbf{\bibinfo{volume}{10}}, \bibinfo{pages}{2207368} (\bibinfo{year}{2023}).

\bibitem[{\citenamefont{Ron et~al.}(2024)\citenamefont{Ron, Crestani, Kux, Liu, Al-Dam, Monzo, Gauthier, S{\'a}ez, and Gov}}]{ron2024emergent}
\bibinfo{author}{\bibfnamefont{J.~E.} \bibnamefont{Ron}}, \bibinfo{author}{\bibfnamefont{M.}~\bibnamefont{Crestani}}, \bibinfo{author}{\bibfnamefont{J.~M.} \bibnamefont{Kux}}, \bibinfo{author}{\bibfnamefont{J.}~\bibnamefont{Liu}}, \bibinfo{author}{\bibfnamefont{N.}~\bibnamefont{Al-Dam}}, \bibinfo{author}{\bibfnamefont{P.}~\bibnamefont{Monzo}}, \bibinfo{author}{\bibfnamefont{N.~C.} \bibnamefont{Gauthier}}, \bibinfo{author}{\bibfnamefont{P.~J.} \bibnamefont{S{\'a}ez}}, \bibnamefont{and} \bibinfo{author}{\bibfnamefont{N.~S.} \bibnamefont{Gov}}, \bibinfo{journal}{Nature Physics} pp. \bibinfo{pages}{1--11} (\bibinfo{year}{2024}).

\end{thebibliography}

\end{document}

\newcommand{\be}{\begin{equation}}
\newcommand{\ee}{\end{equation}}
\newcommand{\bea}{\begin{eqnarray}}
\newcommand{\eea}{\end{eqnarray}}
\newcommand{\nn}{\nonumber}
\renewcommand{\theequation}{S-\arabic{equation}}
\renewcommand{\figurename}{Fig.}
\renewcommand{\thefigure}{S-\arabic{figure}}
\renewcommand{\thetable}{S-\arabic{table}}
\renewcommand{\tablename}{Table}

\title{Supplementary Information - Chemotaxis of branched cells in complex environments}

\author{Jiayi Liu$^{1,2}$, Jonathan E. Ron$^{2}$, Giulia Rinaldi$^{3}$, Ivanna Williantarra$^{3}$, Antonios Georgantzoglou$^{3,4}$, Ingrid de Vries$^{5}$, Michael Sixt$^{5}$, Milka Sarris$^{3}$ and Nir S. Gov$^{2,3}$}
\affiliation{$^{1}$Department of Physics, Yale University, New Haven, CT, USA}
\affiliation{$^{2}$Department of Chemical and Biological Physics, Weizmann Institute of Science, Rehovot, Israel}
\affiliation{$^{3}$Department of Physiology, Development and Neuroscience, Downing Site, University of Cambridge, Cambridge, UK}
\affiliation{$^{4}$Novo Nordisk Foundation Center for Stem Cell Medicine (reNEW), Department of Biomedical Sciences, University of Copenhagen, Denmark}
\affiliation{$^{5}$Institute of Science and Technology Austria (ISTA), Klosterneuburg, Austria}

\maketitle

\section*{S-1. \textit{In-vivo} experimental methods}

\subsection{Zebrafish Husbandry and Preparation}
Tg(lyzC:Gcamp6f) transgenic zebrafish line were maintained adhering to the UK Home Office regulations, UK Animals (Scientific Procedures) Act 1986, which was reviewed by the University Biomedical Service Committee. Adult zebrafish were maintained and bred according to standard protocols \cite{westerfield2007zebrafish}. Embryos were collected at 3 hours post fertilisation, bleached for 5 minutes using 0.003\% NaOCl (Cleanline, CL3013), followed by three times washing in E3 media (as described in \cite{williantarra2024visualising}). Embryos were then raised at 28$^o$C in E3 medium supplemented with 0.003\% PTU (1-phenyl 2-thiourea; Sigma Aldrich, P7629) to inhibit pigmentation and maintain optical transparency. 3 days post-fertilization (dpf) larvae were used on the day of imaging.

\subsection{Bacterial Culture and Preparation}
A day before the imaging day, a single colony of Pseudomonas aeruginosa (strain PAO1 from Martin Welch), isolated via a four-way streak on Pseudomonas Isolation Agar (BD Difco, 292710) supplemented with cetrimide and nalidixic acid (E\&O Laboratories Ltd, LS0006) and glycerol (Fisher Scientific, G/0650/17), was grown in 5 mL of antibiotic-free LB broth (Formedium, LBX0102) at 37$^o$C with shaking for 24 hours. On the imaging day, the overnight culture was diluted and incubated to logarithmic phase (OD$_{600}=0.6–0.8$). Bacterial concentration was adjusted to $3 × 10^5$ CFU/ml using spectrophotometric estimation, based on the assumption that OD$_{600}=0.5$ corresponds to $10^7$ CFU/ml (based on our unpublished data). Bacteria were pelleted and washed twice with PBS (Sigma Aldrich, D8537) to remove residual medium before resuspension in Ringer solution (as described in \cite{poplimont2020neutrophil}) containing 0.16 mg/ml (1x) Tricaine (Merck, A5040).

\subsection{Zebrafish Mounting and Infection}
Transgenic larvae expressing the most prominent calcium indicator, Gcamp6f (cDNA originally described by \cite{chen2013ultrasensitive}), were screened to ensure high expression. Larvae displaying strong fluorescence and healthy morphology were selected for imaging. Anesthetised larvae were mounted in a 1:1 mix of 2\% low-melting agarose (Invitrogen, 16520) and 2× Tricaine in a custom-built coverslip chamber, composed of glass coverslips covering both the top and bottom of the samples and sealed onto a metallic ring. Larvae were oriented laterally and agarose was allowed to solidify. After the agarose was solidified, the tail fin was exposed by carefully removing the surrounding agarose under a dissecting microscope using a tweezer and a capillary needle, allowing bacterial access post-wounding. The chamber was then filled with 1 ml of Ringer solution containing 1x Tricaine and the diluted PAO1 suspension before sealing.

\subsection{Two-Photon Imaging and Laser Wounding}
Laser ablation \cite{poplimont2020neutrophil} and time-lapse imaging were performed using a LaVision TriM Scope multiphoton microscope equipped with an electro-optic modulator for rapid power modulation. A Spectra-Physics Insight DeepSee dual-line laser was tuned to 900 nm for imaging and 1,040 nm for ablation, with the imaging power set to $\sim500$ mW at the specimen plane. Image acquisition was performed using ImSpector Pro software (5.0.284.0, LaVision Biotec, \textcopyright 1998-2016).
Two-photon microscopy was performed using a 25×/NA 1.05 water dipping objective lens, with ddH$_2$O applied to the lens and the correction collar set to 0.17 to match the refractive index. Out of all the mounted larvae, the larvae with the best overall health (normal circulation, heartbeat and morphology) and no damages or aberrant wounding due to mounting was selected for image acquisition. The ventral fin within the caudal hematopoietic tissue of the selected larvae was then centered using brightfield optics. A region of interest with a diameter of 40 $\mu$m was defined as the wound area on a single focal plane (as superficial as possible), scanned at 240 nm/pixel with a 15 $\mu$s dwell time. Z-stacks were acquired every 20 s with a step size of 2 $\mu$m, consisting of ~20 planes. Laser wound was triggered after a 2-minute pre-wound baseline, followed by 3 hours of post-wound imaging. Focus was maintained throughout acquisition. Image stacks were processed in Fiji (ImageJ 1.52p, June 2019, publicly available; \cite{schindelin2012fiji}) using maximum intensity z-projections to generate neutrophil swarming timelapse videos of which were then used for subsequent analysis. 

\subsection{Imaging Analyses}
Analysis of neutrophil trajectories was performed in Imaris v8.2 (Bitplane AG) on 2D maximum intensity projections of the 4D time-lapse movies (\textit{in-vivo} experiments) and 3D time-lapse movies (\textit{in-vitro} experiments) (as previously described in \cite{georgantzoglou2022two,williantarra2024visualising}). Cell trajectories were manually tracked overtime, speed and straightness coefficient of the trajectories were calculated with the imaging software. The projected area of the cells overtime was calculated by manually drawing the perimeter of the cells in Fiji (Image J) and using the measure tool. 

\section*{S-2. \textit{In-vitro} experimental methods for cells migrating in a regular array of pillars}

\begin{figure}[h!]
    \centering
    \includegraphics[width=1\textwidth]{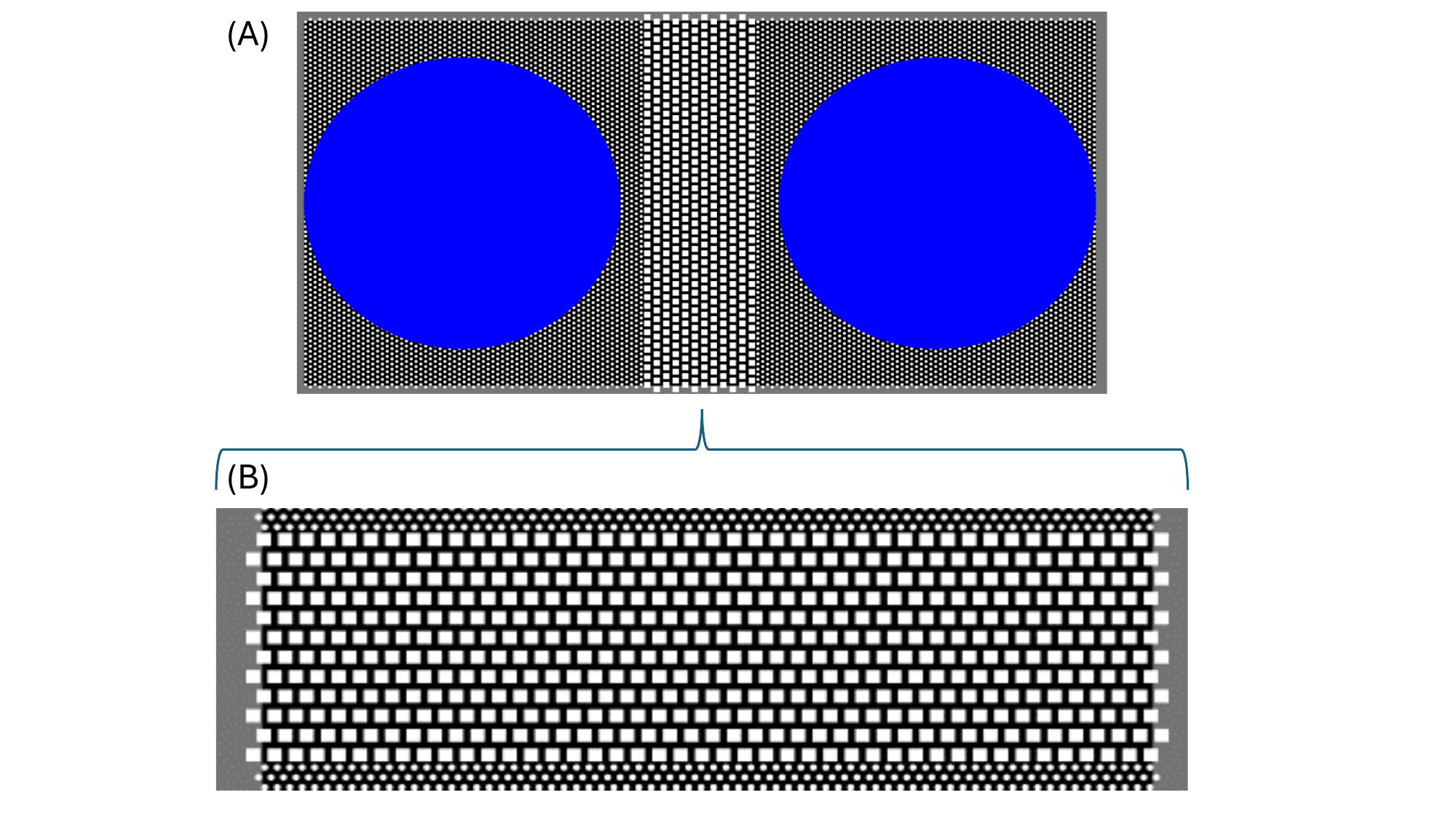}
    \caption{Microfabricated PDMS devices design. (A) Blue dots indicate loading areas for cells and chemoattractant. Loading areas are surrounded by an area of circular pillars, which serve as an antechamber from where cells enter the 0,7 x 2,5mm analysis area. (B) Analysis area harbors 10x10$\mu$m rectangular pillars of 3.8$\mu$m height. Pillars are distanced 5$\mu$m and 3$\mu$m in vertical vs. horizontal direction, and arranged in a hexagonal lattice.}
    \label{figS0}
\end{figure}

We describe here the microfabrication process of the PDMS devices, their design and manufacturing (Fig.\ref{figS0}) \cite{renkawitz2018micro}.

\subsection{Photomask}
The Photomask is chrome patterned on a glass plate and used to fabricate the wafer. Photomask design is drawn in CorelDraw X8 and exported to dxf file format. LinkCad is used to convert dxf to Gerber format. Photomasks are manufactured by https://www.jd-photodata.co.uk/

\subsection{Wafer}
The wafer was backed for 5 min at 110$^o$C, followed by a spin coat Su8 6005 TF (MICROCHEM) and pre-backed for 5-10 at 110$^o$C. The wafer was exposed to 100 mJ/cm$^2$ and post backed at 110$^o$C for 5 min. This was followed by a developing step with SU8 developer and IPA and 135$^o$C baking for 5 min. The height of the device was measured on a profilometer. Finally the wafer was salinized with 1H,1H,2H,2H perfluorooctyltrichlorosilane for 1 h in a sealed vacuum.

\subsection{PDMS devices}
40 ml 1:10 PDMS Sylgard 184 (Dow Corning) was mixed and degassed for 2 min at 2000 rpm and for 2 min at 2200 rpm in a Thinky ARE-250 mixer/defoamer and poured on the wafer in an aluminum mold, degassed in a desiccator and cured for 2 h at 80$^o$C. The devices were cut in small squares and 2,5-mm holes were punched (harris Unicore biopsy puncher) on both sides as loading ports. The devices were cleaned with tape (scotch magic tape) and air blown. Cover slip (\#1,5, 22x22, Menzel) and device were plasma activated for 2 min in a plasma cleaner (Harrick Plasma). The activated side of the device was placed on the charged side of the coverslip and baked  at 95$^o$C for 15 min to achieve bonding. 
Devices attached to the coverslips were glued with paraplast x-tra (sigma) onto the bottom of a 6 cm cell culture dish so that it covers a central hole of 17 mm diameter.  1-2 h before adding cells devices were incubated with R10 media (RPMI 1640 media (21875091 Gibco) supplemented with 10\% FCS (Gibco) and Penicillin-Streptomycin), in a cell culture incubator at 37$^o$C and 5\% CO$_2$. 

\subsection{PLB-985 (promyelocytic leukemia blasts) cells}
PLB-985 cells (promyelocytic leukemia blasts) were obtained from the DSMZ (PLB-985 ACC 139). Cells were grown in R10 media. 3-4 days before the experiment, cells where differentiated by adding 1,25\% of DMSO to the R10 media. Before the experiment cells were stained with 10$\mu$M TAMRA (Invitrogen) diluted in PBS for 5 min in the dark and washed 3x in R10 media. 

Imaging cells in micropillar devices
Liquid was removed from both loading ports. One port was filled with 5-7$\mu$l of fMLP (50 mM) and the other port with 5-7$\mu$l of cell suspension (20.000 cells/$\mu$l). The loaded devices were placed in the incubator for minimum 1 h. Movies of cells migrating in the devices were acquired with an imaging interval of 10 s using a Nikon Ti2E inverted widefield Microscope equipped with a Plan Apo λ 20x/0.75 DIC 1 air PFS objective, a monochrome CMOS sensor camera and a custom-built climate chamber (37$^o$C, 5\% CO2, humidified).

\section*{S-3. Model equations}

\begin{figure}[b]
    \centering
    \includegraphics[width=1\textwidth]{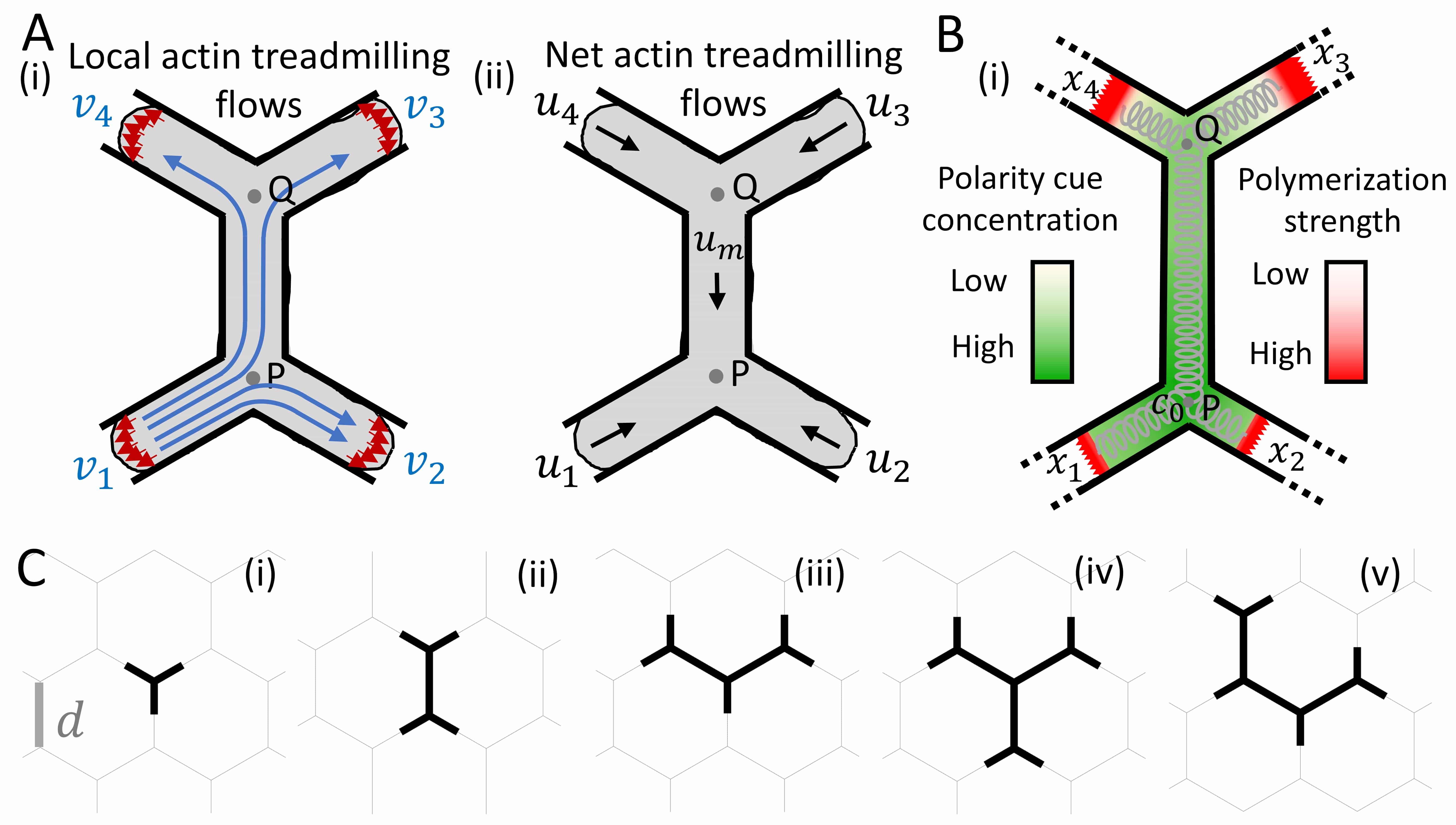}
    \caption{ (A) (i) Illustrations of the local actin treadmilling flows at each edge of the arms, and the split of the local actin flow of arm 1. (ii) Illustrations of the net actin treadmilling flows in each segment of the cell. (B) An example of the concentration field of the polarity cue, which is affected by the advection flows. The cell elasticity is denoted by the grey springs. (C) Shape of the cells that are spanning different number of junctions. (i) One, (ii) two, and (iii) three junctions. (iv-v) Two possible cell shapes for cells spanning four junctions.}
    \label{figS1}
\end{figure}

For a cell with \(N\) \((N \geq 3)\) arms, i.e., spanning across \(N-2\) junctions (Fig.~\ref{figS1}), the dynamics of the arm \(i\) are described by three variables: 1) its length \(x_i\), 2) the fraction of active slip-bonds adhesion \(n_i\) at the leading edge, and 3) the local actin treadmilling flow velocity \(v_i\) at the leading edge. The dynamic equations for these variables are given by \cite{liu2024shape}

\begin{equation} \label{x_i}
{\dot{x}}_i = \frac{1}{\Gamma_i} [v_i - k \, (L-1)]
\end{equation}

\begin{equation} \label{n_i}
{\dot{n}}_i = r \, (1-n_i) - n_i \, \exp{\left[\frac{-v_i+k\,(L-1)}{f_s\,n_i}\right]}
\end{equation}

\begin{equation} \label{v_i}
{\dot{v}}_i = -\delta \, (v_i-v_i^*) +  \sigma \xi_t
\end{equation}

In Eq.\ref{x_i}, \(\Gamma_i\) is a non-constant friction coefficient that depends on the direction of motion of the arm's leading edge, given by

\begin{equation}
\Gamma_i = \Theta(\dot{x}_i )+ (1-\Theta(\dot{x}_i)) \, n_i \, \exp{\left[\frac{v_i-k\,(L-1)}{f_s\,n_i}\right]}
\end{equation}
where \(\Theta\) is a Heaviside function and \(\kappa\) is the effective spring constant of the bond-linkers. When the arm is extending, the friction acts as a constant drag \(\Gamma_i=1\), while when the arm is retracting, the friction is due to the adhesion of the slip-bonds. 

Eq.\ref{x_i} is a simplified description of the protrusive traction forces, which can be further elaborated to include the adhesion dependence of these forces \cite{mukherjee2023actin}. The restoring force of the cell elasticity in Eqs.\ref{x_i},\ref{n_i} is described by a simple spring term, where \(k\) is the effective elasticity of the cell (of rest length 1). 

In Eq.\ref{n_i}, \(f_s\) describes the susceptibility of the slip-bonds to detach due to the applied force, and \(r\) is the effective cell-substrate adhesiveness. 

In Eq.\ref{v_i}, \(\delta\) is the rate at which the local actin flows relax to the steady-state solutions \(v_i^*\), which is given by
\begin{equation} \label{v_ss}
v_i^* = \beta \frac{c_s}{c_s+c_i(x_i)}
\end{equation}
where $c_i(x_i)$ is the concentration of actin polymerization inhibitor at the tip of arm $i$ (at the coordinate $x_i$), $c_s$ is the saturation concentration, and \(\beta\) is the maximal actin polymerization speed at the arm edges. The term \(\sigma \xi_t\) in Eq.\ref{v_i} describes the noise in the actin polymerization activity, with a random Gaussian form of amplitude \(\sigma\).

The calculation of the spatial distribution of the inhibitor concentration along the different branches $c_i(x_i)$ was performed. The spatial distribution is composed of exponential sections, maintaining continuity at the junctions, a no-flux boundary condition, and a constant total amount of inhibitor within the cell.

The total length of the cell is given by 
\begin{equation}
L=\sum_{i} x_i+(N-3)\,d 
\end{equation}
where \(d\) is the distance between two adjacent junctions on the network. Following the initialization of \(x_i\), \(n_i\) and \(v_i\) of each arm, their temporal evolution during the migration process can be obtained through numerical integration of Eq.\ref{x_i}, Eq.\ref{n_i} and Eq.\ref{v_i}. In this study, we employed the symmetric initial condition.

\begin{table}[htbp]
    \centering
    \begin{tabular}{|c|c|}
    \hline
    Parameter & Value \\
    \hline
    $c$ & 3.85 \\
    $D$ & 3.85 \\
    $k$ & 0.8 \\
    $f_s$ & 5 \\
    $r$ & 5 \\
    $\kappa$ & 20 \\
    $\delta$ & 250 \\
    \hline
    \end{tabular}
    \caption{Some of the model parameters used in this study.}
    \label{tabS1}
\end{table}

\begin{figure}[b]
    \centering
    \includegraphics[width=1\textwidth]{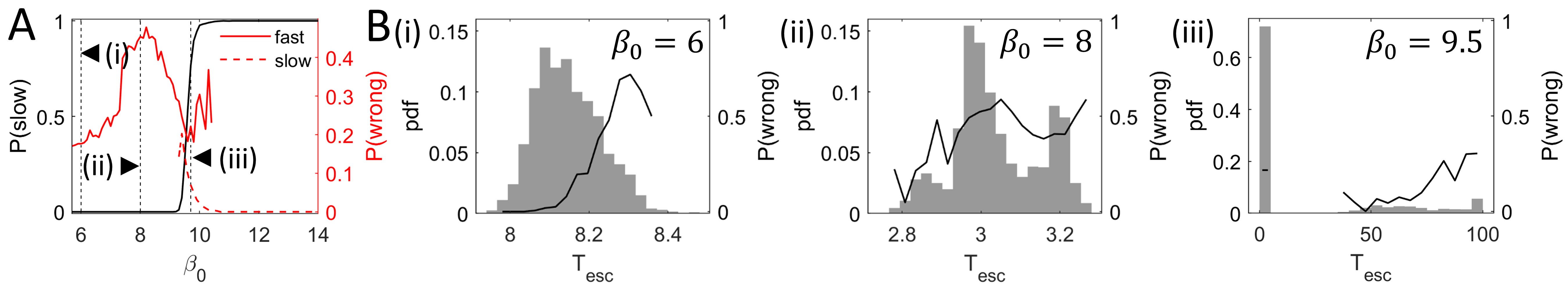}
    \caption{ (A) For the system in main text Fig.2B(i), we plotted the mean probability of the cell getting into the "slow-mode" on the junction \(P(slow)\) (black line). We plotted separately the accuracies of the fast and slow processes, by red solid and dashed lines respectively. (i-iii) Distributions of \(T_{esc}\) (histograms) and \(P(wrong)\) (solid line) as functions of \(T_{esc}\), for the values of \(\beta_0\) indicated by the vertical dashed lines in (A). Key parameters: \(\epsilon=0.001\), \(\sigma=0.1\). }
    \label{figS2}
\end{figure}

\section*{S-4. Effect of noise and hexagon size on chemotactic migration}
\begin{figure}[htbp]
    \centering
    \includegraphics[width=1\textwidth]{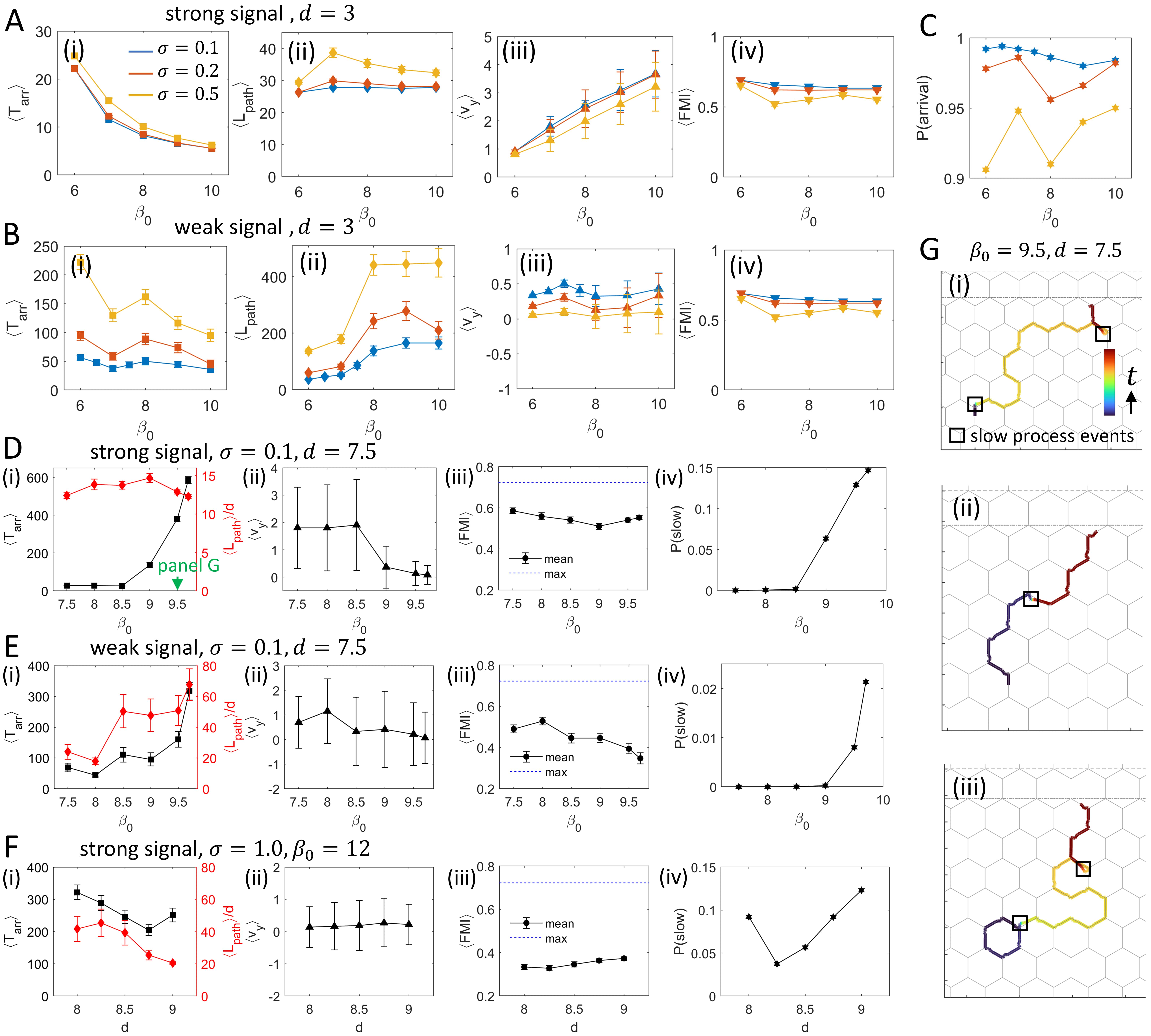}
    \caption{Effect of cellular internal noise \(\sigma\) and hexagonal edge size $d$ on chemotactic migration. (A-B) On small grids (\(d=3\)): (i) \(\langle T_{\text{arr}} \rangle\), (ii) \(\langle L_{\text{path}} \rangle\), (iii) \(\langle v_y \rangle\), and (iv) \(\langle FMI \rangle\) as functions of \(\beta_0\). (A) and (B) correspond to the strong signal regime and the weak signal regime, respectively. (C) Probability of arriving at the source at \(T_{max}=1000\), for the weak signal regime of (B). (D-E) On large grids (\(d = 7.5\)): (i) \(\langle T_{\text{arr}} \rangle\) and \(\langle L_{\text{path}} \rangle\), (ii) \(\langle v_y \rangle\), (iii) \(\langle FMI \rangle\) (black solid line) and the maximal theoretical \(FMI\) (blue dashed line), and (iv) \(P(\text{slow})\) as functions of \(\beta_0\). (D) and (E) correspond to the strong signal regime and the weak signal regime, respectively. (F) Effect of grid size \(d\) on: (i) \(\langle T_{\text{arr}} \rangle\) and \(\langle L_{\text{path}} \rangle\), (ii) \(\langle v_y \rangle\), (iii) \(\langle FMI \rangle\), and (iv) \(P(\text{slow})\) for \(\beta_0 = 12\), in the strong signal regime (as in (A)) and under a large noise level (\(\sigma = 1.0\)). (G) Typical trajectories of the cell's C.O.M., for \(\beta_0=9.5\), \(d=7.5\), and \(\sigma=0.1\). Black squares mark the occurrence of slow process events along the trajectory. Maximal simulation time: \(T=1000\). Other key parameter: \(\epsilon=0.2\).}
    \label{figS3}
\end{figure}

We investigated the effect of cellular internal noise on the efficiency of chemotactic migration. As shown in Fig.~\ref{figS3}A-C, for both the strong and weak signal regimes (Fig.~\ref{figS3}A and B, respectively), an increase in noise leads to an increase in arrival time and path length, as well as a decrease in \(y\)-direction velocity and Forward Migration Index (FMI), indicating a reduction in migration efficiency.

In Fig.~\ref{figS3}B, for large $\beta_0$, we see a saturation of the mean path length. This might indicate that the accuracy of cellular DDM has a saturation, such that for $\beta_0>8$ there is no significant further increase in decision-making errors. This is also consistent with the mean arrival time which continues to decrease as $\beta_0$ increases, since the path length remains similar, but the migration speed increases with $\beta_0$.

In Fig.~\ref{figS3}C, we calculated the probability of the cell arriving at the target within the maximal simulation time \(T=1000\), \(P({arrival})\), as the function of $\beta_0$, for the weak signal regime. It can be observed that for larger noises, there are peaks in \(P({arrival})\) at the minima of \(\langle T_{arr} \rangle\) from Fig.~\ref{figS3}B(i), while the remaining feature is the minimum in \(P(arrival)\) around $\beta_0\sim8$.

In Fig.\ref{figS3}D,E, we investigated the dependence of mean \(T_{\text{arr}}\), \(L_{\text{path}}\), \(v_y\), and FMI on \(\beta_0\) for larger grid size \(d=7.5\). For large \(\beta_0\), cells exhibit slow-mode events during migration, with their probability, \(P(slow)\), increasing significantly as \(\beta_0\) increases (Fig.~\ref{figS3}D,E(iv)). This leads to a substantial increase in \(T_{\text{arr}}\).

In the strong signal regime (Fig.~\ref{figS3}D), when \(\beta_0\) reaches the threshold where slow events occur, a further increase in \(\beta_0\) results in a significant rise in \(T_{\text{arr}}\), while \(L_{\text{path}}/d\) decreases and FMI increases, indicating an improvement in accuracy. In the weak signal regime (Fig.~\ref{figS3}E), the occurrence of slow events also leads to a significant increase in \(T_{\text{arr}}\), but the increase in \(L_{\text{path}}/d\) and the decrease in \(FMI\) suggest that, in this case, the slow mode does not enhance accuracy.

In Fig.~\ref{figS3}F, we examined the effect of grid size (\(d\)) on cell migration in the strong signal regime under large \(\beta_0=12\) and high \(\sigma=1\) conditions. We observed that for larger \(d\), the quantities \(P_{\text{slow}}\) and \(FMI\) increase, while \(L_{\text{path}}/d\) decreases (Fig.~\ref{figS3}F(i,iii,iv)), indicating that the slow mode indeed enhances migration accuracy, consistent with the case of strong signal and small noise (Fig.~\ref{figS3}C). Note that for the cells that do not arrive at the target within the \(T_{\text{max}}\), we assigned \(T_{\text{arr}}= 1000\).

In Fig.~\ref{figS3}G we plotted typical trajectories of the cell's C.O.M., for \(\beta_0=9.5\), \(d=7.5\), and \(\sigma=0.1\). Black squares mark the occurrence of slow process events along the trajectory. Note that these slow-mode events are of a type that is different from those found on a single junction (as shown in Fig.2) and on the hexagonal network in Fig.5F. Instead of the long arms forming in the direction of migration, we find that the cell undergoes a stick-slip event, after which one of the two long arms form along the direction from which the cell arrived at the junction (as shown in Fig.\ref{figS8}).

\section*{S-5. Detailed analysis of the dynamics of the cell's trajectory}

\begin{figure}[htbp]
    \centering
    \includegraphics[width=1\textwidth]{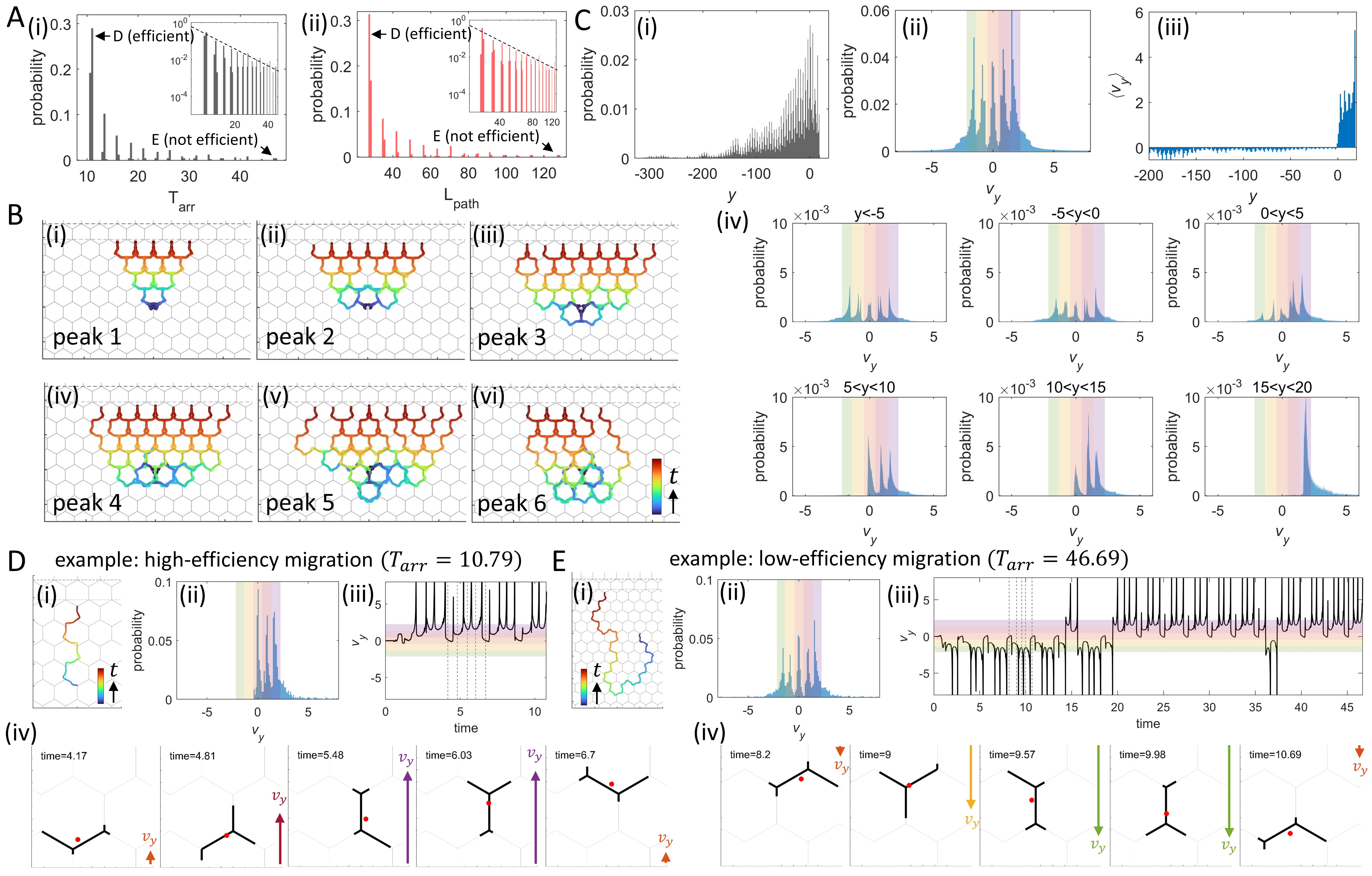}
    \caption{Analysis of the \(T_{arr}\), \(L_{path}\), \(y\) and \(v_y\) in the slow regime (\(C/c_0=0.01\)). A) Distributions of (i) \(T_{arr}\)  and (ii) \(L_{path}\). B) Trajectories corresponding to the first six peaks of the \(T_{arr}\) distribution and the \(L_{path}\) distribution in A. C) Distributions of (i) \(y\) and (ii) \(v_y\) of the cell's centroid. (iii) The \(v_y\) distributions at different \(y\) ranges. (iv) The \(\langle v_y \rangle\) as a function of \(y\). D) Example of high efficiency migration. (i) Trajectory of cell's C.O.M.. (ii) Distribution of \(v_y\) during the trajectory. (iii) Time series of \(v_y\) during the trajectory. (iv) Snapshots for the time stamps (gray dashed lines) in (iii). Key parameters: \(\epsilon=0.2, d=3, \beta_0=8, \sigma=0.1\).}
    \label{figS4}
\end{figure}

We investigated the dynamics of the cell's trajectory in more detail for the weak signal regime. We plotted the distributions of \(T_{arr}\) and \(L_{path}\) (Fig.~\ref{figS4}A), using \(\beta = 8\) as an example. Both distributions exhibit discrete peaks, with a constant interval between adjacent peaks. By plotting the trajectories corresponding to the first six peaks of \(T_{\text{arr}}\) and \(L_{\text{path}}\) (Fig.~\ref{figS4}B) we conclude that the constant interval between the arrival time (and path length) peaks arises from the additional time (or path length) that is added to the trajectory each time the cell makes an additional incorrect decision along the path, i.e. a turn that takes it away from the chemokine source.

We also plotted the distribution of \(y\) and \(v_y\) (Fig.~\ref{figS4}C(i-ii)), where the peaks result from the periodic migration patterns of the cell as it moves along the hexagonal edges. Furthermore, we plotted \(\langle v_y \rangle\) (Fig.~\ref{figS4}C(iii)) and the \(v_y\) distributions within specific ranges of \(y\) (Fig.~\ref{figS4}C(iv)). As expected, \(\langle v_y \rangle\) is positive only at positive \(y\)-positions, where the cell is sufficiently close to the chemokine source. Further away from the source the cell loses the very weak signal and performs isotropic diffusion.

For illustration, we provide examples of efficient and inefficient migration paths (corresponding to the 1st and 15th peaks in Fig.~\ref{figS4}A), shown in Fig.~\ref{figS4}D,E. Panel (i) displays the trajectory, while (ii) shows the \(v_y\) distribution. Panel (iii) presents the time series of \(v_y\), with shaded regions corresponding to the respective \(v_y\) ranges indicated by the colors in (ii). We found that the \(v_y\) time series exhibits a highly periodic pattern, as the cell performs stick-slip migration over the regular hexagonal network. This migration pattern, with fast slip events, is further demonstrated in panels (iv), where we present typical snapshots of the cell shape and C.O.M. at the times indicated by the vertical black dashed lines in panels (iii).

\section*{S-6. Comparing the experimental response of the cell to the chemokine with the model}

\begin{figure}[htbp]
    \centering
    \includegraphics[width=1\textwidth]{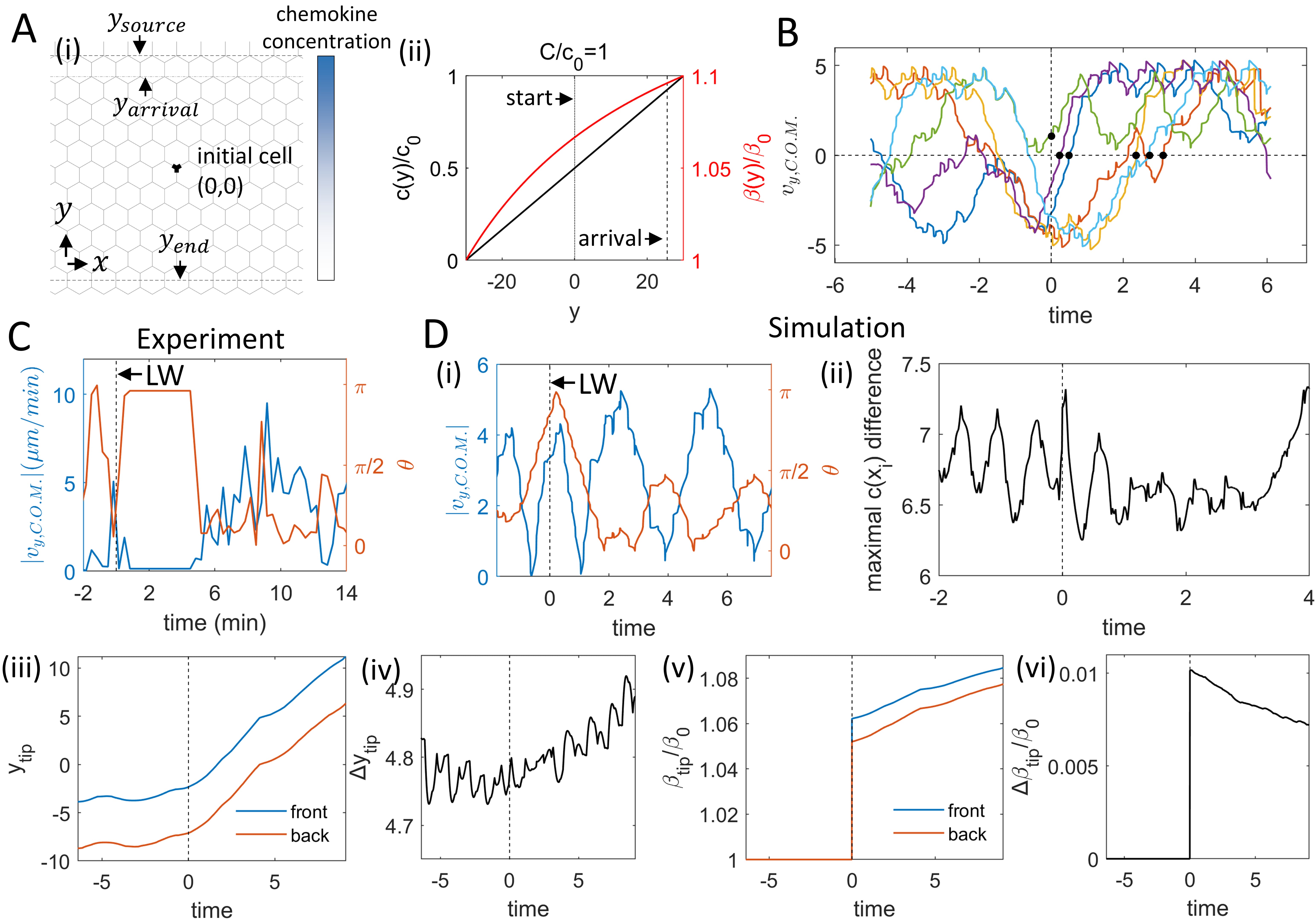}
    \caption{Multiple-junction model with a chemokine line source that decays linearly. (A) (i) Schematic representation of the model. Two gray dashed lines indicate the position of the chemokine source and the position where the source decays to zero, respectively. The gray dash-dotted line marks the arrival position. (ii) Normalized chemokine concentration, \(c(y)/c_0\) (black line), and the enhancement of actin polymerization speed, \(\beta(y)/\beta_0\), as functions of \(y\). (B) Dynamics of \( v_{y, \text{C.O.M.}} \) for six randomly selected simulation trajectories. The black dots indicate the BM time corresponding to each trajectory. (C) Dynamics of the C.O.M. velocity in \(y\)-direction, \(|v_{y,\text{C.O.M.}}|\), of the experimental cell (main text Fig.~4). (D) Dynamics of the simulated cell's (i) C.O.M. velocity in the \(y\)-direction, \(|v_{y,\text{C.O.M.}}|\), (ii) maximal difference between the polarity cue concentration at the arm tips, \(c(x_i)\), (iii) \(y\)-coordinates of the front tip (maximum \(y\)-coordinate) and back tip (minimum \(y\)-coordinate), (iv) difference in \(y\)-coordinates between the front and back tips, (v) \(\beta\) enhancement, \(\beta/\beta_0\) at the front and back tips, (vi) difference in \(\beta\) enhancement between the front and back tips. Black dashed lines mark the LW time. Parameters: \(\epsilon=0.1, d=3, \beta_0=12, \sigma=0.5\).}
    \label{figS5}
\end{figure}

To simplify the theoretical model, we assumed a spatially linear variation of the chemokine concentration for the comparison with the experimental data (Fig.~\ref{figS5}A(i)). As in the previous theoretical analysis , the concentration is directed along the \(y\)-axis. It reaches its maximum value \(c_0\) at the source position \(y_{\text{source}}\). Now, we assumed that at a specific \(y\) position, \(y_{\text{end}}\), the concentration drops to zero. The concentration varies linearly between \(y_{\text{source}}\) and \(y_{\text{end}}\), meaning that the chemokine concentration function \(c(y)\) in our model (Eq.3) takes the following linear form:

\begin{equation}
c(y) = c_0 \frac{y - y_{\text{end}}}{y_{\text{source}} - y_{\text{end}}}
\end{equation}

We set \(y_{\text{source}} = -y_{\text{end}} = 8d\). At the start of each simulation, the cell is symmetrically positioned at the origin \((x, y) = (0, 0)\). The chemokine profile parameters are set as \(\epsilon = 0.1\) and \(C/c_0 = 1\). 

Under this setting, the normalized chemokine concentration and the enhancement of actin polymerization speed as functions of \(y\) are presented in Fig.~\ref{figS5}A(ii). It can be observed that although the chosen parameter \(C/c_0 = 1\) corresponds to the weak signal ("slow saturation") regime, the linear variation of \(c(y)\) ensures that as soon as the chemical cue is introduced, the cell perceives a strong level of signal, provided it is within the region between \(y_{\text{source}}\) and \(y_{\text{end}}\) (which is indeed satisfied in our setup).

The "Beginning of movement" (BM) time (Fig.~4D) is defined in our simulations as the time point at which cell movement in the direction of the chemical gradient begins (Fig.~\ref{figS5}B). In \cite{georgantzoglou2022two} this time point was defined similarly, where cell motion was first identified as initiated towards the LW signal. It is determined individually for each cell based on the time point at which the cell initiates movement in the "wound" (\(y\)) direction. 

The effect of the chemokine on the FMI of the cells is given in Fig.~5E for the two durations used in \cite{georgantzoglou2022two}. It is clear that the significant improvement in the FMI due to the external gradient, in these short time windows, only appears for large $\beta_0$, which is the regime that fits the experimental observations for the WT cells.

In Fig.~\ref{figS5}C,D(i), we show the dynamics of the cell's C.O.M. velocity in the \(y\)-direction, \(|v_{y,\text{C.O.M.}}|\), and the angle \(\theta\) between the C.O.M. velocity vector and the positive \(y\)-axis for the experiment and the simulation (see schematic in Fig.~4A,B). In both cases, the cells slow down following the introduction of the chemokine gradient, then re-polarize and increase their speed towards the source. 

The origin of deceleration in the model can be demonstrated by plotting the maximal front-back difference in the polarity cue along the cell length (Fig.~\ref{figS5}D(ii)), and by plotting the polarity concentrations at the tips of the cellular arms (Fig.~4B). Prior to the introduction of the chemokine gradient, the polarity cue difference oscillates as the cell periodically changes its length and number of arms spanning the junction while moving on the hexagonal network. At time$=0$, the cell was polarized such that its leading edge was pointing away from the source. This means that the back of the cell is exposed at time $t=0$ to a higher chemokine concentration than the front, and therefore the actin polymerization is more enhanced at the cell's rear than at its front. A transient enhancement of the actin polymerization activity at the side of the cell facing the chemokine gradient was also observed in neutrophils \cite{georgantzoglou2022two}. The difference in polymerization activity at the cells edges due to a different chemokine concentration results in a smaller overall front-back retrograde flow and weaker advection of the polarity cue. The front-back polarity cue difference therefore decreases (Fig.~\ref{figS5}D(ii)), which slows down the cell migration. By reducing the cell polarity the protrusive activity becomes more uniform among all the cellular protrusions, not limited to the cell front, and facilitates the rotation of the cell. As the cell re-establishes its front-back polarity in the direction of the chemokine source (the front-back polarity cue difference recovers its large values, Fig.~\ref{figS5}D(ii)), and the cell resumes its fast migration. Note that the oscillations in the simulated cell speed are much more regular than in the experiment (Fig.~\ref{figS5}C,D(i)), but the angle of migration makes oscillations of similar size in both cases.

In Fig.~\ref{figS5}D(iii), we give the dynamics of the arm tips with the largest (front) and smallest (back) $y$-values from the experiment and the simulation used in Fig.~4. The length of the cell along the $y$-axis is shown in Fig.~\ref{figS5}D(iv), and we find that the cell is stretched by the chemokine gradient in the gradient direction.

The chemokine-induced enhancement of the actin polymerization activity at the front and back tips of the cell (as defined by their $y$-coordinate in (i)), is shown in Fig.~\ref{figS5}D(v). This difference (Fig.~\ref{figS5}D(vi)) is slowly decreasing over time, as the cell moves up the gradient and the overall chemokine concentration around the cell increases.

\section*{S-7. Comparing the migration characteristics of experimental cells and the model}

\begin{figure}[h!]
    \centering
    \includegraphics[width=1\textwidth]{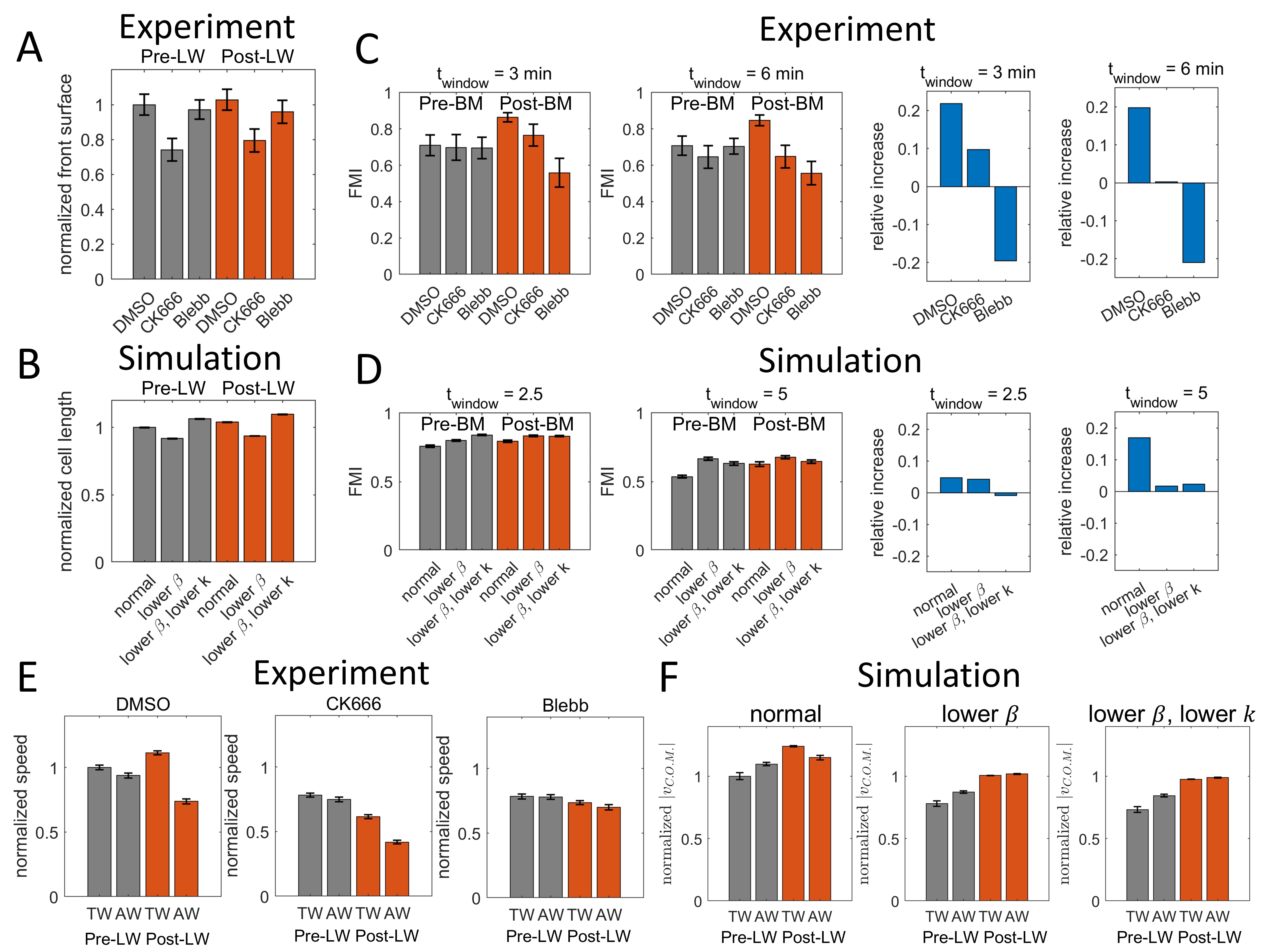}
    \caption{Comparison of the model with the chemotaxis properties of Neutrophiles \textit{in-vivo} \cite{georgantzoglou2022two}. (A) Normalized front surface of the experimental cell (normalized by the DMSO value before BM), before and after the BM time (grey and red respectively). (B) Normalized cell length of the simulated cell (normalized by the normal cell value before BM) before and after the BM time (grey and red respectively). (C) FMI of the experimental cell's C.O.M. under different drug treatments within a specific time window (left two panels) and the relative increase after the BM time compared to before the BM time (right two panels). (D) FMI of the simulated cell's C.O.M. under different parameter settings (corresponding to the respective drug treatments in experiments) within a specific time window (left two panels) and the relative increase after the BM time compared to before the BM time (right two panels). (E) Normalized speed of the experimental cell's C.O.M. (normalized by the DMSO value TW(pre)), towards (TW) and away (AW) from the wound, before and after the LW time (grey and red respectively). (F) Normalized speed of the simulated cell's C.O.M. (normalized by the normal cell value TW(pre)), before and after the BM time (grey and red respectively). For wild-type (normal) cells, CK-666-treated cells, and blebbistatin-treated cells, the corresponding simulation parameters were chosen as: \((\beta,k)=(12.0,0.8)\), \((\beta,k)=(10.0,0.8)\), and \((\beta,k)=(10.0,0.7)\), respectively. Other key parameters: \(\epsilon=0.1, d=3, \sigma=0.5\).}
    \label{figS6}
\end{figure}

The total cell length is observed to be significantly diminished by decreasing actin polymerization activity (CK666, Fig.\ref{figS6}A)\cite{georgantzoglou2022two}. Compared to the WT cells, we fit a reduced $\beta_0$, by $\sim17\%$ in the CK666-affected cells (Fig.~\ref{figS6}A)\cite{georgantzoglou2022two}.

The effect of blebbistatin, which inhibits myosin-II contractility, is described in our model as a decrease in the cell contractile-stiffness parameter $k$ and the retrograde flow of actin $\beta_0$ \cite{ron2024emergent}. While decreasing $k$ acts to make the cell longer, the lower $\beta_0$ decreases the forces that act to elongate the cell, so that combined the cell is similar in length to the WT, as observed in the experiments (Fig.~\ref{figS6}A,B)\cite{georgantzoglou2022two}.

In Fig.~\ref{figS6}C,D, we compare the experimental \cite{georgantzoglou2022two} and simulated changes to the FMI for the WT (DMSO) and drug-treated cells. The relative changes are qualitatively captures by the model, especially the increase in FMI due to the chemokine for the WT cells, and the vanishing of this effect upon drug application. This, again, points to the WT cells residing in the high-$\beta_0$ regime of our model, which is the only regime where we see that the FMI is significantly increased by the chemokine gradient.

Similarly in Fig.~\ref{figS6}E,F the normalized cell speed is compared between the WT and drug-treated cells \cite{georgantzoglou2022two}. In the WT cells a large difference is found between the cell speed towards and against the chemokine direction, while this is very much reduced for the drug-treated cells, which also have lower overall migration speed compared to the WT (Fig.~\ref{figS6}E). These qualitative features are captured by the model (Fig.~\ref{figS6}F).

%Our conclusion from the comparisons above, that the neutrophiles correspond to the high-activity cells of our model is further supported by the following features of their chemotaxis (Fig.~S-5). We show here two trajectories of neutrophiles migrating towards the laser-wound in the zebrafish larva fin. When the neutrophiles start their trajectory close to the wound, they tend to move in direct paths, as shown. When starting from further away, we often find that cells exhibit highly meandering paths (Fig.~\ref{fig5}). Such meandering paths are found in our model only for the high-$\beta_0$ regime (see Fig.~\ref{fig3}D). Furthermore, we can often identify that the cells display the slow-mode behavior (Fig.~\ref{fig2}C(iii)), as shown in Fig.~\ref{fig5}.

\begin{figure}[h!]
    \centering
    \includegraphics[width=0.7\textwidth]{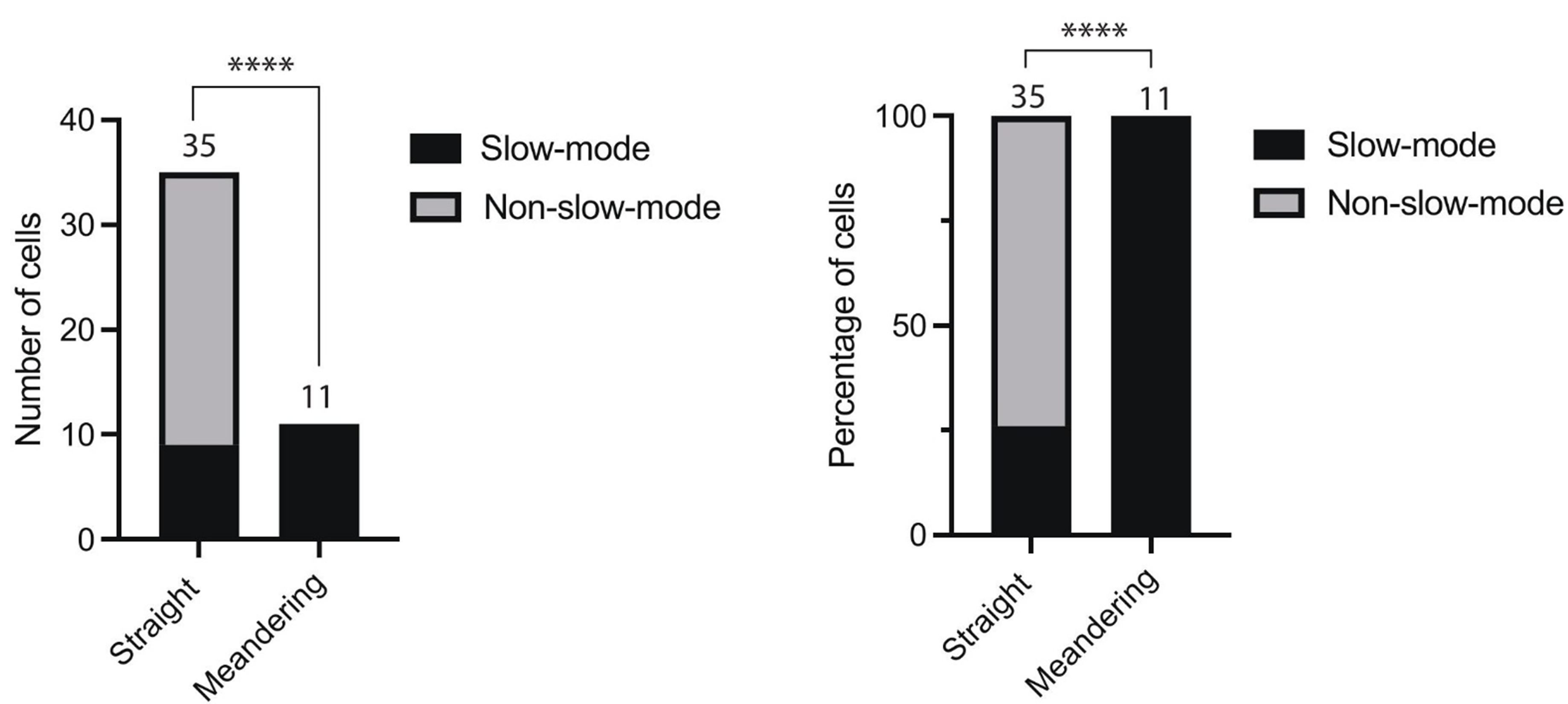}
    \caption{Characterizing "slow-mode" events for cells \textit{in-vivo} (as in Fig.~5A-C). GCamp6F transgenic zebrafish embryos (3 dpf) were infected with \textit{Pseudomonas aeruginosa}, laser-wounded, and imaged using a two-photon confocal microscope, as previously described. Neutrophil trajectories were determined manually using the imaging software IMARIS. Meandering and straight trajectories were classified based on the straightness coefficient quantified by the software (\(> 0.60\) considered as straight). Trajectories were also categorized based on the presence or absence of slow-mode events. The graph shows both the percentage and absolute count of cells displaying slow-mode events within the straight and meandering categories. Percentages are based on 39 and 11 cells, respectively, as indicated in the graph. Videos with an overall recruitment greater than 15 cells were selected. \(N = 46\) neutrophil trajectories were quantified from 7 independent videos/embryos. \(\ast\ast\ast\ast\, P < 0.0001\), Chi-square test (and Fisher’s exact test).}
    \label{figS7}
\end{figure}

\begin{figure}[h!]
    \centering
    \includegraphics[width=1\textwidth]{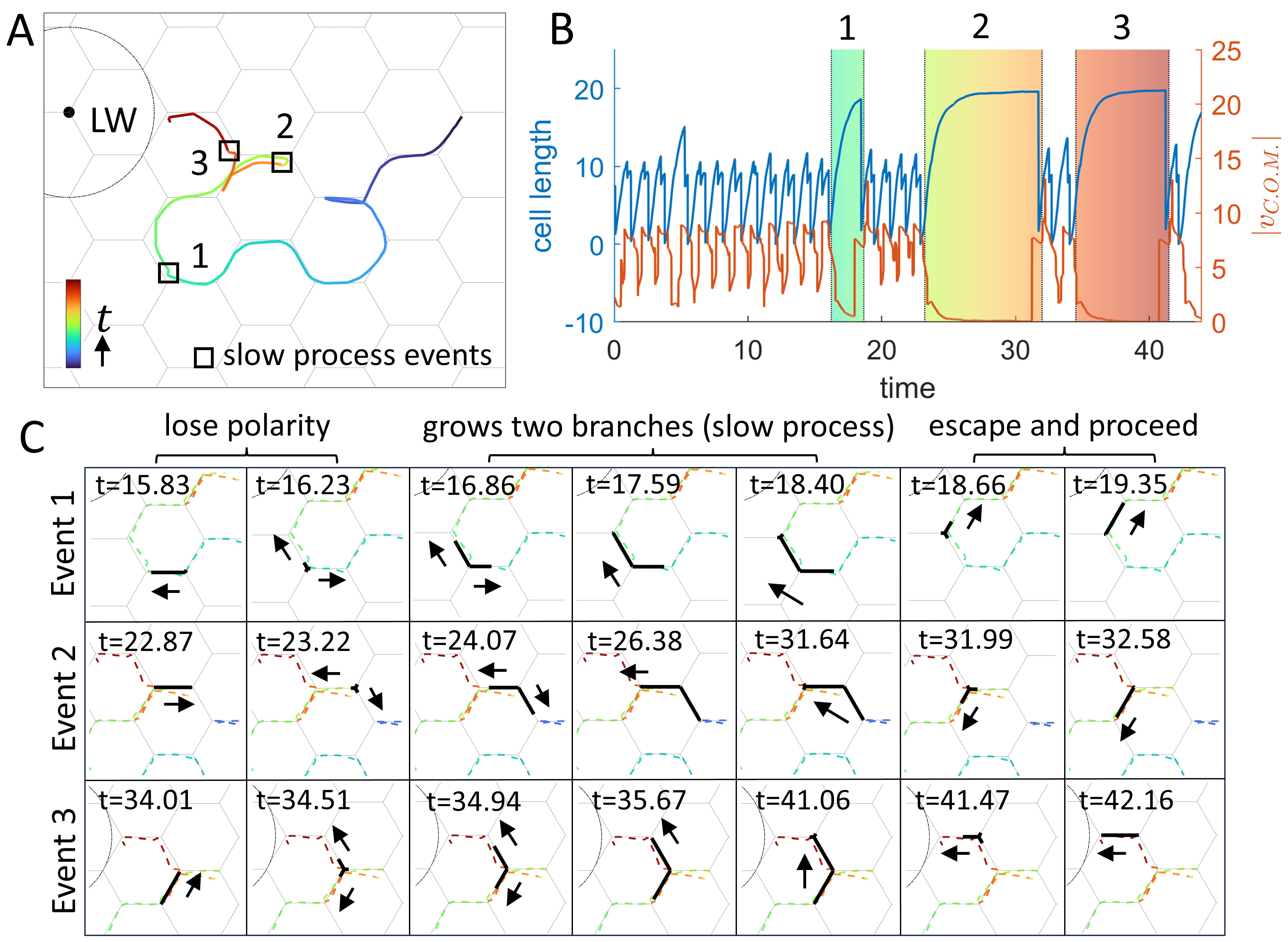}
    \caption{Illustration of the second type of slow mode during cell migration on the large grids. (A) A representative simulation trajectory of a cell migrating toward the LW, with the trajectory color indicating migration time. Black boxes highlight time periods along the trajectory in which slow-mode events occur, corresponding to the images in (C). (B) Dynamics of the cell length and the C.O.M. speed of the cell during migration. The two colored regions correspond to the three slow-mode events marked in (D). (C) Simulation snapshots of the cell during the two slow-mode events. Parameters: \(C/c_0=0.01, \epsilon=0.2, d=9, \beta_0=12, \sigma=1.2\).}
    \label{figS8}
\end{figure}

\bibliography{bibliography}